\documentstyle[preprint,aps,graphicx]{revtex}
\tightenlines
\begin{document}
 
\title{The Finite Temperature $SU(2)$ Savvidy Model with a Non-trivial
Polyakov Loop}
 
\author{Peter N. Meisinger and Michael C. Ogilvie}
 
\address{ Dept. of Physics
Washington University
St. Louis, MO 63130}
 
\date{\today}
 
\maketitle
 
\begin{abstract}   

We calculate the complete one-loop effective potential
for $SU(2)$ gauge bosons at temperature $T$ as a
function of two variables:
$\phi $, the angle associated with a non-trivial Polyakov loop,
and $H$, a constant background chromomagnetic field.
These two variables are indicators for confinement and
scale symmetry breaking, respectively.
Using techniques broadly applicable to finite temperature field theories,
we develop both low and high temperature expansions.
At low temperatures,
the real part of the effective potential $V_R$
indicates a rich phase structure,
with a discontinuous
alternation between confined $(\phi=\pi)$
and deconfined phases $(\phi=0)$. 
The background field
$H$ moves slowly upward from its zero-temperature value
as $T$ increases,
in such a way that $\sqrt{gH}/\pi T$ is approximately
an integer.
This behavior stops at $T_c = 0.722(1) \mu_0$, where $\mu_0$
is a zero-temperature renormalization group invariant scale;
beyond this temperature, the deconfined phase is always preferred.
At high temperatures, where perturbation theory should be
reliable as a consequence of asymptotic freedom,
the deconfined phase  $(\phi=0)$ is always preferred, and
$\sqrt{gH}$ is of order $g^2(T)T$.
The imaginary part of the effective potential $V_I$,
which originates in a tachyonic mode associated with the lowest
Landau level,
is non-zero
at the global minimum of $V_R$ for all temperatures.
A non-perturbative
magnetic screening mass of the form
$M_m = c g^2(T) T$ with a sufficiently large coefficient $c$
removes this instability at high temperature,
leading to a
stable high-temperature phase with
$\phi = 0$ and $H=0$, characteristic of a weakly-interacting gas of
gauge particles. 
The value of $M_m$ obtained is comparable with lattice estimates.

\end{abstract}

\section{Introduction}

\vspace{1pt}One of the most important features of $SU(N)$ gauge theories at
finite temperature is the existence of a deconfinement phase transition. The
essential features of the transition have been well established by lattice
simulations \cite{Boyd:1996bx}.
Below the deconfinement temperature $T_{d}$, the pressure is
essentially zero, because glueball masses are large compared to $T_{d}$.
Above $T_{d}$, the pressure rises, appearing to slowly approach the
blackbody result for a free gas of gauge bosons. 

Within the Euclidean finite temperature formalism, 
the deconfinement transition can be
viewed as the spontaneous breaking of a global symmetry associated with the
center $Z(N)\,$of the gauge group $SU(N)$ \cite{Svetitsky:1982gs}.
In this
formalism, the temporal direction is periodic, with period $%
\beta =1/T$. 
The Polyakov loop, defined as
the Euclidean time-ordered exponential 
\begin{equation}
{\mathcal P}\left( \overrightarrow{x}\right) ={\mathcal T}\,\exp
\left[ ig\oint d\tau
\,A_{0}\left( \overrightarrow{x},\tau \right) \right] 
\end{equation}
is the natural order parameter for the deconfinement transition. 
We will regard ${\mathcal P}\,$as
an abstract element of the group $SU(N)$, and use $Tr_{R}{\mathcal P}$
to denote its
trace in the representation $R$. Below the deconfinement transition
temperature $T_{d}$, in the confining phase, the $Z(N)$ symmetry is
unbroken, which in turn implies that the thermal average of the fundamental
representation trace vanishes,
$\left\langle Tr_{F}{\mathcal P}\right\rangle =0$. Above 
$T_{d}$, the $Z(N)$ symmetry is spontaneously broken, and $\left\langle
Tr_{F}{\mathcal P}\right\rangle \neq 0$. Calculation of the effective
potential for ${\mathcal P}$ in perturbation theory indicates that the $Z(N)$ symmetry is broken
at high temperature, but gives no information about its restoration at low
temperature \cite{Gross:1981br,Weiss:1981rj,Weiss:1982ev}.

A constant chromomagnetic field is the simplest non-trivial field
configuration for which the one-loop functional determinant can
be evaluated analytically. This is essentially 
the Saviddy model \cite{Matinian:1976mp,Savvidy:1977as}.
The Savvidy model 
is an interesting laboratory for perturbative
calculations in non-Abelian gauge theories.
The essential feature of this model is
the perturbative prediction that the vacuum of a
non-Abelian gauge theory has a
chromomagnetic condensate.
As pointed out by Savvidy, this behavior can be
inferred from asymptotic freedom. 
A direct path to this result is a calculation of the zero-temperature
effective potential $V(T=0)$ for constant non-Abelian magnetic
fields, which shows a non-trivial minimum.
However,
as first discussed by Nielsen and Olesen \cite{Nielsen:1978rm},
the zero-temperature effective potential has
a tachyonic instability,
{\it i.e.}, an instability with respect to long
wavelength fluctuations.
This gives $ V(T=0, H \neq 0)\,$a negative imaginary component,
indicating that a constant field must decay towards the true
vacuum state, which is unknown.
Nevertheless, the non-trivial minimum of $V$ at $H \neq 0$
is often
regarded as evidence for the dynamic breaking of scale invariance in gauge
theories, and for the existence of a gauge field condensate.

The Savvidy model at finite temperature allows us to
examine the coupling of the Polyakov loop to a gauge field condensate.
\vspace{1pt}Lattice results indicate that the expectation values of other
key observables are coupled to the Polyakov loop. For example, at the
deconfinement transition in pure $SU(3)$ gauge theory, which is first order,
both the Polyakov loop and the plaquette expectation values are
discontinuous. In fact, the plaquette expectation values are related to the
internal energy density and pressure via the trace anomaly, and thus must
jump at a first order transition. A simple strong coupling expansion reveals
that plaquette expectation values depend on Polyakov loops via topologically
nontrivial strong coupling diagrams that wind around the lattice in the
Euclidean time direction. A similar behavior is seen for the chiral
condensate \cite{Meisinger:1995qr,Meisinger:1995ih,Meisinger:ea}
Thus we expect to find that the field $H$
couples to the Polyakov loop in the Savvidy model. 

We calculate the effective potential at all temperatures, including
the effect of a non-trivial Polyakov loop as well as a constant
non-Abelian magnetic field. From this effective potential, we
find that the Savvidy model distinguishes between low-temperature behavior
and high-temperature behavior in a novel manner, 
At low temperatures, the Saviddy model exhibits an intriguing oscillation
between confined and deconfined phases as the temperature is varied. 
We regard this as important, because
understanding the mechanism which determines $\left\langle
Tr_{F}{\mathcal P}\right\rangle $ as $T$ is varied in the full theory
is tantamount to understanding the
deconfinement transition, and very likely the origin of confinement.
The high-temperature behavior of the Savvidy model is
similar to that of a free gas of gauge bosons, with a trivial
Polyakov loop.

Previous work on the Savvidy model at high temperatures 
with a trivial Polyakov loop has shown that the tachyonic instability found
at zero temperature persists at high temperatures
\cite{Ninomiya:eq}.
This is potentially much more serious than instability at zero or low
temperatures, because asymptotic freedom is generally taken to imply
the utility of perturbative calculations in high-temperature gauge theories.
The persistence of the tachyonic instability at high temperature
is a barrier to the use of perturbation theory in this regime.
Remarkably, a non-trivial Polyakov loop can counteract the tachyon
instability at finite temperature
\cite{Meisinger:1997jt,Starinets:vi}.
We will explore this mechanism in detail
in what follows. However, the basic point is simple: a nontrivial Polyakov
loop acts as an additional positive mass term for gauge field fluctuations,
and removes the tachyonic instability in some circumstances.
Ultimately this approach fails: 
at one loop, the imaginary part of the effective potential
is non-zero at the global minimum of $V$ for all temperatures.
An attractive possibility for high temperatures is that
a non-perturbative magnetic mass stabilizes the model
at one loop.
As we will demonstrate below, a sufficiently large magnetic mass
will not only remove the tachyonic stability, but also leads
to a straightforward characterization of the high-temperature
behavior with $H=0$ as the global minimum of $V$.

In the process of performing the evaluation of the effective potential, we
have collected many elementary methods for doing one loop finite temperature
calculations. Many of these are well known. However, we have largely avoided
the use of hypergeometric functions
\cite{Haber1:1982,Haber2:1982} and
finite temperature zeta-function techniques \cite{Actor:1986zf,Actor:1987cf},
which are powerful but somewhat opaque. The alternative high temperature
techniques we use may be of interest, independent of the particular features
of the Savvidy model, especially since they are widely applicable, and
retain the periodic properties associated with the Polyakov loop 
\cite{Meisinger:2001fi}.
At appropriate points, we also stress the physical interpretation of key
results, including high-$T\,$and low-$T$ expansions.

The next section develops the formalism required
for the evaluation of the one loop effective potential. Section III provides
the details of the low temperature expansion, expanding on the results of
\cite{Meisinger:1997jt}.
A high temperature expansion is derived in section
IV, and section V examines these results in the context of dimensional
reduction. Section VI considers the stability of $H\neq 0\,$states at high
temperatures. In section VII we present our conclusions. There are three
technical appendices.

\section{The Effective Potential}

In this section, we introduce the formalism necessary for the
evaluation of the one-loop effective potential for $SU(2)$
gauge bosons at finite temperature. This is accomplished by calculating the
partition function in background gauge, with the background field providing
both a constant non-Abelian magnetic field and a non-trivial Polyakov loop.
In order to carry out the calculation, we choose the color magnetic field
and the Polyakov loop to be simultaneously diagonal. We take the color
magnetic field $H$ to point in the $x_{3}$ direction. The external vector
potential can be chosen to be

\begin{equation}
A_{2}=Hx_{1}{\frac{\tau _{3}}{2}} 
\end{equation}
\vspace{1pt}which gives rise to a chromomagnetic field 
\begin{equation}
F_{12}=H{\frac{\tau _{3}}{2}}\text{.} 
\end{equation}
The Polyakov loop is specified by a constant $A_{0}$ field, given in the
fundamental representation by 
\begin{equation}
A_{0}={\frac{\phi }{{g\beta }}}\frac{\tau _{3}}{2} 
\end{equation}
where the range of $\phi $ can be taken as $0\leq \phi \leq 2\pi $. The
trace of the Polyakov loop is given by 
\begin{equation}
Tr_{F}({\mathcal P})=2\cos (\phi /2) 
\end{equation}
in the fundamental representation and by 
\begin{equation}
Tr_{A}({\mathcal P})=1+2\cos (\phi ) 
\end{equation}
in the adjoint representation. 
In general, the $N-1$ eigenvalues of the $SU(N)$ Polyakov loop
are not determined by the fundamental representation trace
\cite{Meisinger:2001cq}.
However, 
for $SU(2)$ and $SU(3)$, the trace of the Polyakov loop in the fundamental
representation does determine the eigenvalues.
In the case of $SU(2)$, the global $Z(2)$ symmetry takes
$\phi$ into $2\pi-\phi$. Unless the $Z(2)$ symmetry is spontaneously
broken, the variable $\phi$ must have the value $\pi$,
corresponding to $Tr_{F}({\mathcal P})=0$.

At tree level in the loop expansion, the
effective potential is given by 
\begin{equation}
V^{(0)}=\frac{1}{2}H^{2}\text{,} 
\end{equation}
the classical field energy. As explained in \cite{Nielsen:1978rm},
the
external field $H$ gives rise to Landau levels in the gluon functional
determinant. The one-loop contribution to the free energy has the form 
\cite{Nielsen:1978rm,Ninomiya:eq};
\begin{eqnarray}
V^{(1)} &=&V_{0}^{(1)}+V_{\pm }^{(1)}  \nonumber \\
&=&\sum_{n}\,{\frac{1}{\beta }}\,\int \,{\frac{d^{3}\vec{k}}{(2\pi )^{3}}}%
\,\ln \left( {\omega }_{n}^{2}+\vec{k}^{2}\right) + \nonumber\\
&&{\frac{1}{2}}\,\sum_{m=0}^{\infty }\,\sum_{n,\pm }\,{\frac{1}{\beta }}\,\,{%
\frac{gH}{2\pi }}\,\int \,{\frac{dk_{3}}{2\pi }}\,\ln \left[ \left( \omega
_{n}-{\frac{\phi }{\beta }}\right) ^{2}+2gH\left( m+{\frac{1}{2}}\pm
1\right) +k_{3}^{2}\right] +  \label{e2.1} \nonumber\\
&&{\frac{1}{2}}\,\sum_{m=0}^{\infty }\,\sum_{n,\pm }\,{\frac{1}{\beta }}\,\,{%
\frac{gH}{2\pi }}\,\int \,{\frac{dk_{3}}{2\pi }}\,\ln \left[ \left( \omega
_{n}+{\frac{\phi }{\beta }}\right) ^{2}+2gH\left( m+{\frac{1}{2}}\pm
1\right) +k_{3}^{2}\right]
\label{eq:V1}
\end{eqnarray}
where the $\omega _{n}=2\pi n/\beta $ are the usual Matsubara frequencies,
and the sum over $n\,$is over all integer values. The sum over $m\,$is the
sum over Landau levels. The contribution $V_{0}^{(1)}$ comes from the first
term, due to the neutral $Z^{0}$; the second and third terms
give $V_{\pm }^{(1)}$, arising from the
charged $W^{+}$ and $W^{-}$ gauge bosons.
The
terms $2gH(m+1/2\pm 1)$ are the allowed Landau levels for the charged gauge
fields. 
The $\pm 2gH$ term originates in the familiar coupling of spin to
an external magnetic field. In this case, the $g$ factor is $2$,
and $S_{z}=\pm 1$.

For $m=0$ and $k_{3}$ sufficiently small, the negative sign
gives rise to tachyonic modes which are responsible for destabilizing the
original Savvidy vacuum, as first pointed out by Nielson and Oleson
\cite{Nielsen:1978rm}.
As discussed in \cite{Meisinger:1997jt,Starinets:vi},
it is possible to avoid tachyonic
contributions provided $Tr_{F}({\mathcal P})$ is sufficiently small in
magnitude. The worst behavior occurs in the $n=0\,$Matsubara mode and the $%
m=0\,$Landau level. The determinant will be strictly real provided 
\begin{equation}
\beta \sqrt{gH}<\phi <2\pi -\beta \sqrt{gH}\text{.} 
\label{e2.3}
\end{equation}
Thus, it is possible that the Savvidy vacuum, or a similar field
configuration with apparent tachyonic modes, 
could be stabilized in the confining
phase by the non-trivial Polyakov loop. As a practical matter, the vanishing
of the imaginary part of $V\,\ $for this range of parameters provides us
with an important check for various expressions.

With the aid of the standard product representation \cite{GandR} 
\begin{equation}
\frac{\cosh (x)-\cos (a)}{1-\cos (a)}=\prod_{k=-\infty }^{\infty }\left[
1+\left( \frac{x}{2\pi k+a}\right) ^{2}\right] 
\end{equation}
we can write the one-loop effective potential in the form: 
\begin{eqnarray}
V^{(1)} &=&2{\frac{1}{2}}\,\sum_{m=0}^{\infty }\sum_{\pm }\,\,{\frac{gH}{%
2\pi \beta }}\,\int \,{\frac{dk}{2\pi }}\,\ln \left\{ \cosh \left[ \beta {%
\omega }_{\pm }(m,k)\right] -\cos (\phi )\right\}  \nonumber \\
&+&{\frac{1}{\beta }}\,\int \,{\frac{d^{3}\vec{k}}{(2\pi )^{3}}}\,\ln
\left\{ \cosh \left[ \beta \varepsilon _{k}\right] -1\right\} \text{,}
\label{e2.7}
\end{eqnarray}
where $\varepsilon _{k}=|\vec{k}|$ and the variables ${\omega }_{\pm }$ are
defined by 
\begin{equation}
{\omega }_{\pm }^{2}(m,k)=2gH\left( m+{\frac{1}{2}}\pm 1\right) +k^{2}\text{.%
} 
\end{equation}
Note that $\omega _{-}^{2}\,\ $can be negative. In order to obtain the
correct sign for the imaginary part of $V$, it is necessary to supply the
standard Feynman $i\varepsilon $ prescription as needed, ${\omega }_{\pm
}^{2}\rightarrow {\omega }_{\pm }^{2}-i\varepsilon $, with $\varepsilon
\rightarrow 0$. The contribution of the $Z^{0}$ can be written as 
\begin{equation}
V_{0}^{(1)}=\int \,{\frac{d^{3}\vec{k}}{(2\pi )^{3}}}\,\left[ \varepsilon
_{k}+\frac{2}{\beta }\ln \left( 1-e^{-\beta \varepsilon _{k}}\right) \right] 
\end{equation}
The first term is an irrelevant vacuum energy contribution, and the second
term is the free energy of a massless, free gauge boson. Similarly, we can
write $V_{\pm }^{(1)}\,$as 
\begin{equation}
V_{\pm }^{(1)}=\sum_{m=0}^{\infty }\sum_{\pm }\,{\frac{gH}{2\pi }}\int %
{\frac{dk}{2\pi }}\,\left[ \omega _{\pm }+\frac{1}{\beta }\ln \left(
1-e^{-\beta \omega _{\pm }+i\theta }\right) +\frac{1}{\beta }\ln \left(
1-e^{-\beta \omega _{\pm }-i\theta }\right) \right] 
\end{equation}
\vspace{1pt}

There is a clear separation of the zero-temperature part of the
effective potential and the finite temperature part. We write the finite
temperature part of $V_{0}^{(1)}$ and $V_{\pm }^{(1)}$ as $U_{0}$ and $%
U_{\pm }$, respectively. All ultraviolet divergences reside in the zero
temperature part 
\begin{eqnarray}
V^{(1)}(T =0)&=&V_{0}^{(1)}(T=0)+V_{\pm }^{(1)}(T=0) \nonumber\\
&=&\int {\frac{d^{3}\vec{k}}{(2\pi )^{3}}}\,\varepsilon
_{k}+\sum_{m=0}^{\infty }\sum_{\pm }\,\,{\frac{gH}{2\pi }}\,\int \,{\frac{dk%
}{2\pi }}\,{\omega }_{\pm }(m,k)\text{.}
\end{eqnarray}
The finite, $H$-dependent part of $V^{(1)}(T=0)\,$was first derived by
Nielson and Oleson 
\cite{Nielsen:1978rm}; for the reader's convenience, we review their
derivation in Appendix 1. With
an appropriate choice of renormalization constant, $V^{(0)}$ can be absorbed
into $V^{(1)}(T=0)$ and $V(T=0)$\ is given by 
\begin{eqnarray}
V(T =0)&=&V^{(0)}+V^{(1)}(T=0) \nonumber\\
&=&\frac{11g^{2}H^{2}}{48\pi ^{2}}\,\,\ln \left( {\frac{gH}{{\mu }_{0}^{2}}}%
\right) -i\frac{(gH)^{2}}{8\pi ^{2}}\text{.}
\end{eqnarray}
where $\mu _{0}$ is a renormalization group-invariant parameter that sets
the scale for the gauge theory. The minimum of the effective potential
occurs at $gH=\Lambda _{S}^{2}$, where $\Lambda _{S}=\mu _{0}\exp \left(
-1/4\right) $. Note that the combination $gH$ is renormalization group
invariant in background field gauge, and henceforth $g\,\,$and $H\,\,$will
generally appear together.

\section{Low Temperature Expansion}

The one-loop finite temperature contribution to $V\,$\ from the $Z^{0}$
gauge boson, which we write as $U_{0}$, is given by

\begin{equation}
U_{0}=\frac{2}{\beta }\int \,{\frac{d^{3}\vec{k}}{(2\pi )^{3}}}\,\ln \left(
1-e^{-\beta \varepsilon _{k}}\right) =-2\frac{\pi ^{2}}{90\beta ^{4}}\text{.}
\end{equation}
This is the free energy density of a boson gas with two spin degrees of
freedom. The phase $\phi $ does not appear because the $Z^{0}$ is
charge-neutral. Because the free energy is extensive in the volume, this is
also the negative of the $Z^{0}\,$contribution to the pressure. The $W^{\pm
}\,$contribution $U_{\pm }$ is somewhat more difficult to evaluate.
Expanding the logarithm, the contribution of the Landau levels to the
effective potential can be written in the form: 
\begin{equation}
U_{\pm }=-{\frac{2}{\beta }}\sum_{m=0}^{\infty }\sum_{\pm }\,\,{\frac{gH}{%
2\pi }}\,\int \,{\frac{dk}{2\pi }}\,\left[ \,\sum_{n=1}^{\infty }\,{\frac{%
\cos (n{\phi })}{n}}e^{-n\beta {\omega }_{\pm }(m,k)}\right] \text{.} 
\end{equation}
This expression has a natural interpretation in terms of path integrals. At
finite temperature, there are particle trajectories which wind around
space-time in the Euclidean temporal direction, and are thus topologically
non-trivial. For such trajectories, there is a factor of $\exp \left( in\phi
\right) $ when the net winding number is $n$. Alternatively, we can consider 
$i\phi $ as a chemical potential continued to imaginary values. 

Some care must be exercised, because ${\omega }_{-}(0,k)=\sqrt{%
2gH\left( 0+{\frac{1}{2}}-1\right) +k^{2}}\,$is imaginary for $k\,$
sufficiently small. The contribution to $U_{\pm }$ from $\omega _{-}$ with $%
m=0\,$ is 
\begin{equation}
u(-,0)=-{\frac{2}{\beta }}\,\,{\frac{gH}{2\pi }}\,\sum_{n=1}^{\infty }\,{%
\frac{\cos (n{\phi })}{n}}\int \,{\frac{dk}{2\pi }}\,\,e^{-n\beta {\omega }%
_{-}(0,k)}\text{.} 
\end{equation}
We define $k_{c}^{2}=gH\,$\ so that ${\omega }_{-}(0,k_{c})=0$. Then $u(-,0)$
can be written as 
\begin{equation}
u(-,0)=-{\frac{2gH}{\beta \pi }}\,\sum_{n=1}^{\infty }\,{\frac{\cos (n{\phi }%
)}{n}}\left[ \int_{0}^{k_{c}}\,{\frac{dk}{2\pi }}\,\,e^{-n\beta {\omega }%
_{-}(0,k)}+\int_{k_{c}}^{\infty }\,{\frac{dk}{2\pi }}\,\,e^{-n\beta {\omega }%
_{-}(0,k)}\right] \text{.} 
\end{equation}
The analytic continuation of the first integral is carried out using the
general prescription that $-i\varepsilon$
is added to $2gH\left( n+{\frac{1}{2}}\pm 1\right)$,
analogous to the standard $%
m^{2}-i\varepsilon $ prescription. 
For the case of the $(-,0)$ contribution, this means that 
$gH\rightarrow gH+i\varepsilon $.
The first integral can be written as 
\begin{equation}
\int_{0}^{k_{c}}\,{\frac{dk}{2\pi }}\,\,e^{-n\beta \sqrt{k^{2}-k_{c}^{2}-i%
\varepsilon }}=k_{c}\int_{0}^{\pi /2}\,{\frac{d\theta }{2\pi }\cos \theta }%
\,\,\left[ \cos \left( n\beta k_{c}\cos \theta \right) \right] +\frac{k_{c}}{%
4\pi }\pi iJ_{1}(n\beta k_{c})\text{.} 
\end{equation}
while the second integral is 
\cite{AandS}
\begin{equation}
\int_{k_{c}}^{\infty }\,{\frac{dk}{2\pi }}\,\,e^{-n\beta \sqrt{%
k^{2}-k_{c}^{2}}}=-\frac{k_{c}}{4}Y_{1}\left( n\beta k_{c}\right) -\frac{%
k_{c}}{2\pi }\int_{0}^{\pi /2}d\theta \,\cos \left( n\beta k_{c}\cos \theta
\right) \cos \theta \text{.} 
\end{equation}
When the two terms are added, the remaining integrals cancel, yielding the
result

\begin{equation}
u(-,0)={\frac{\left( gH\right) }{2\pi \beta }}^{3/2}\,\sum_{n=1}^{\infty }\,{%
\frac{\cos (n{\phi })}{n}}\left[ Y_{1}\left( n\beta k_{c}\right)
-iJ_{1}(n\beta \sqrt{gH})\right] \text{.} 
\end{equation}
The other terms all have the form 
\begin{equation}
u(\pm ,m)=-{\frac{2}{\beta }}\,\,{\frac{gH}{2\pi }}\,\sum_{n=1}^{\infty }\,{%
\frac{\cos (n{\phi })}{n}}\int \,{\frac{dk}{2\pi }}\,\,\exp \left[ -n\beta 
\sqrt{2gH\left( m+{\frac{1}{2}}\pm 1\right) +k^{2}}\right] 
\end{equation}
where $m\geq 1\,$for $u(-,m)$. Using the integral representation 
\begin{equation}
K_{1}(n\beta M)=\frac{1}{2M}\int_{-\infty }^{\infty }dk\,e^{-n\beta \sqrt{%
k^{2}+M^{2}}} 
\end{equation}
we obtain 
\begin{equation}
u(\pm ,m)=-\,\,\frac{gH}{\pi ^{2}\beta }\sqrt{2gH\left( m+{\frac{1}{2}}\pm
1\right) }\,\sum_{n=1}^{\infty }\,{\frac{\cos (n{\phi })}{n}}\,\,K_{1}\left[
n\beta \sqrt{2gH\left( m+{\frac{1}{2}}\pm 1\right) }\right] \text{.} 
\end{equation}
The contribution of each $u(+,m)$ is doubled by the $u(-,m+1)$ term, except $%
u(-,1)$ stands alone, giving a contribution to $U_{\pm }\,$of the form 
\begin{equation}
-\,\,\frac{\left( gH\right) ^{3/2}}{\pi ^{2}\beta }\sum_{n=1}^{\infty }\,{%
\frac{\cos (n{\phi })}{n}}\left[ K_{1}(n\beta \sqrt{gH})+2\sum_{m=0}^{\infty
}\sqrt{2m+3}K_{1}(n\beta \sqrt{\left( 2m+3\right) gH})\right] 
\end{equation}
Combining the results for $V(T=0)$, $U_{0}$, and $U_{\pm }$, we obtain
finally a renormalized effective potential with real component 
\begin{eqnarray}
V_{R} &=&\frac{11(gH)^{2}}{48\pi ^{2}}\,\ln \left( {\frac{gH}{{\mu }_{0}^{2}}%
}\right) -{\frac{2\pi ^{2}}{90\beta ^{4}}} \nonumber\\
&&-{\frac{(gH)^{3/2}}{{\pi }%
^{2}\beta }}\,\sum_{n=1}^{\infty }\,{\frac{\cos (n\phi )}{n}}\left[
K_{1}(n\beta \sqrt{gH})-{\frac{\pi }{2}}Y_{1}(n\beta \sqrt{gH})\right] 
\nonumber \\
&\phantom{=}&-{\frac{2(gH)^{3/2}}{{\pi }^{2}\beta }}\,\sum_{n=1}^{\infty }\,{%
\frac{\cos (n\phi )}{n}}\,\sum_{m=0}^{\infty }\,\sqrt{2m+3}K_{1}\lbrack
n\beta \sqrt{(2m+3)gH}\rbrack  \label{e2.11}
\end{eqnarray}
and imaginary component 
\begin{equation}
V_{I}=-{\frac{\left( gH\right) ^{2}}{8\pi }}-{\frac{\left( gH\right) ^{3/2}}{%
2\pi \beta }}\,\sum_{n=1}^{\infty }\,{\frac{\cos (n\phi )}{n}}\,J_{1}(n\beta 
\sqrt{gH})\text{.} 
\end{equation}

Note that $V_{R}=0$ when $H=0$ and $T=0$. Numerical testing
verifies that $V_{I}$ is indeed zero whenever $\beta \sqrt{gH%
}<\phi <2\pi -\beta \sqrt{gH}$ as required from the expression in terms of
Matsubara frequencies. A similar calculation of $V_{R}\,$and $V_{I}$ has
also been performed by Starinets, Vshivtsev, and Zhukovskii 
\cite{Starinets:vi}.
They also noted the condition for stability, Eq.~(\ref{e2.3}). In their
work, it appears that the Bessel functions $J_{1}$ and $Y_{1}$ were
inadvertently interchanged in the formulae for the real and imaginary part
of the potential, but otherwise our formulae are identical. Our results are
in numerical agreement with the exact result for $V_{I}$ derived by Cabo,
Kalashnikov, and Shabad \cite{Cabo:1980js}
for the case $\phi =0$, and with a
generalization of their expression for $\phi \neq 0$, derived below.

\vspace{1pt}At low temperatures such that $\beta ^{2}gH\ll 1$, the dominant
contribution to $V_{R}$ comes from the terms involving $Y_{1}$, which arise
from the tachyonic mode, and are entirely responsible for the
temperature-dependent part of $V_{I}$. Thus, we may write at low
temperatures 
\begin{equation}
U_{\pm }\approx \,\,2{\frac{gH}{2\pi }}\,\int_{0}^{\sqrt{gH}}\,{\frac{dk%
}{2\pi }}\,\left[ \frac{1}{\beta }\ln \left( 1-e^{i\beta \sqrt{gH-k^{2}}%
+i\phi }\right) +\frac{1}{\beta }\ln \left( 1-e^{i\beta \sqrt{gH-k^{2}}%
-i\phi }\right) \right] \text{.} 
\end{equation}
The real part is 
\begin{eqnarray}
{Re}U_{\pm }\approx \,\,{\frac{gH}{2\pi }}\frac{1}{\beta }%
\,\int_{0}^{\sqrt{gH}}\,{\frac{dk}{2\pi }}\,\Biggl[ &&\ln \left( 2-2\cos \left(
\beta \sqrt{gH-k^{2}}+\phi \right) \right) \nonumber\\
&&+\ln \left( 2-2\cos \left( \beta 
\sqrt{gH-k^{2}}-\phi \right) \right) \Biggr],
\end{eqnarray}
and the deriviative of $Re U_{\pm }$\ with respect to $\phi $ is 
\begin{eqnarray}
\frac{\partial }{\partial \phi }Re U_{\pm }\approx \,\,{\frac{gH}{%
2\pi }}\frac{1}{\beta }\,\int_{0}^{\sqrt{gH}}\,{\frac{dk}{2\pi }}\,\Biggl[ 
&&\frac{\sin \left( \beta \sqrt{gH-k^{2}}+\phi \right) }{1-\cos \left( \beta 
\sqrt{gH-k^{2}}+\phi \right) }\nonumber\\
&&-\frac{\sin \left( \beta \sqrt{gH-k^{2}}-\phi
\right) }{1-\cos \left( \beta \sqrt{gH-k^{2}}-\phi \right) }\Biggr] \text{.} 
\end{eqnarray}
It immediately follows that the minimum of $U_{\pm }$ must occur near either 
$\phi =0$ or $\phi =\pi $ for low temperatures. Careful numerical analysis
of the low temperature formulae confirms that the global minimum
indeed is given by either $\phi =0$ or $\phi =\pi $. 

A similar
analysis of the derivative with respect to $gH$ shows that the minima of $%
Re U_{\pm }$ are found at $\beta \sqrt{gH}\approx 2n\pi $ when $%
\phi =0$, and $\beta \sqrt{gH}\approx \left( 2n+1\right) \pi $ for $%
\phi =\pi $. Numerical analysis shows that $\phi =0$ is the global minimum
of $V_{R}$ for all temperatures $T>0.722(1)\mu _{0}$. As the temperarature
is lowered, the global minimum of $V_{R}$ alternates between $\phi =0$ and $%
\phi =\pi $. For the range of temperatures $0.198(1)\mu _{0}<T<0.722(1)\mu
_{0}$, the global minimum is at $\phi =\pi $, $\beta \sqrt{gH}\approx
\pi $. As the temperature is lowered towards zero, the minimum of $gH\,$%
moves toward its zero-temperature value, and the global minimum of $V_{R}$
continues to alternate between $\phi =0$ and $\phi =\pi $, with
corresponding changes in $gH$. 
In figures 1 through 4 we plot $V_R$ as a function of $\sqrt{gH}/\pi T$
for $\phi=0$ and $\phi=\pi$ for successively lower temperatures,
showing the alternation of the minima. Note how the minima
occur at integer values of the variable  $\sqrt{gH}/\pi T$.
Figure 5 shows how a rapid oscillation is
superimposed on the zero-temperature, $\phi$-independent 
behavior of the $V_R$ at very low temperatures.
Numerical computation of $V_{I\,}$
shows that $V_{I}\neq 0$ at the global minimum for any given temperature.
Thus the Savvidy
model is unstable at low temperatures. However, it is striking that the
low-temperature behavior of the model shows a strong role for the Polyakov
loop, with many local minima of $V_{R}$.

\section{\protect\vspace{1pt}High Temperature Expansion}

In order to develop a suitable high temperature expansion for $V$%
, we return to the original expression for $V_{\pm }^{(1)}$, using the
standard device of Schwinger to express the logarithms of equation 
(\ref{eq:V1}) as an integral 
\begin{equation}
V_{\pm }^{(1)}=-\sum_{m=0}^{\infty }\sum_{n,\pm }\frac{1}{\beta }\frac{gH}{%
2\pi }\int \frac{dk}{2\pi }\int_{0}^{\infty }\frac{dt}{t}\exp \left[
-t\left[ \left( \frac{2\pi n-\phi }{\beta }\right) ^{2}+2gH\left( m+\frac{1}{%
2}\pm 1\right) +k^{2}\right] \right]
\end{equation}
where we have used the symmetry of the expressions under $%
n\longleftrightarrow -n\,$to combine the contributions from $W^{+}$ and $%
W^{-}$. We apply a $\theta _{4}$ $\,$identity of the form

\begin{equation}
\sum_{n}\exp \left[ -\frac{t}{\beta ^{2}}\left( \phi -2\pi n\right)
^{2}\right] =\frac{\beta }{\sqrt{4\pi t}}\sum_{p}\exp \left[ -\frac{\beta
^{2}p^{2}}{4t}+ip\phi \right] 
\end{equation}
and perform the trivial integration over $k\,\,$to obtain 
\begin{equation}
V_{\pm }^{(1)}=-\sum_{m=0}^{\infty }\sum_{\pm }\sum_{p}\frac{gH}{8\pi ^{2}}%
\int_{0}^{\infty }\frac{dt}{t^{2}}\exp \left[ -t\cdot 2gH\left( m+\frac{1}{2}%
\pm 1\right) \right] \exp \left[ -\frac{\beta ^{2}p^{2}}{4t}+ip\phi \right] 
\end{equation}
In the zero temperature limit, only the $\phi$-independent
$p=0$ term survives.
We henceforth omit this term, since it 
is included in $V^{(1)}(T=0)$, evaluated in Appendix A. 
Using the symmetry under $%
p\longleftrightarrow -p$, the remainder of the above expression is the
finite temperature contribution 
\begin{equation}
U_{\pm }=-\sum_{m=0}^{\infty }\sum_{\pm }\sum_{p=1}^{\infty }\frac{gH}{4\pi
^{2}}\int_{0}^{\infty }\frac{dt}{t^{2}}\exp \left[ -t\cdot 2gH\left( m+\frac{%
1}{2}\pm 1\right) \right] \exp \left[ -\frac{\beta ^{2}p^{2}}{4t}\right]
\cos (p\phi )\text{.} 
\end{equation}
We perform the sum over $m$ and $\pm $, obtaining
\begin{eqnarray}
U_{\pm } &=&-\frac{gH}{4\pi ^{2}}\int_{0}^{\infty }\frac{dt}{t^{2}}\left[
e^{-tgH}\coth (tgH)\right] \sum_{p=1}^{\infty }\exp \left[ -\frac{\beta
^{2}p^{2}}{4t}\right] \cos (p\phi ) \nonumber\\
&&-\frac{gH}{4\pi ^{2}}\int_{0}^{\infty }\frac{dt}{t^{2}}e^{tgH}\sum_{p=1}^{%
\infty }\exp \left[ -\frac{\beta ^{2}p^{2}}{4t}\right] \cos (p\phi )
\end{eqnarray}
where the second term is due to the unstable mode. This integral must be
defined by analytic continuation and is responsible for the imaginary part
of $U_{\pm }$. After the shifts $t\rightarrow t/gH\,$for the first term and $%
t\rightarrow -t/gH\,$for the second, we have 
\begin{eqnarray}
U_{\pm } &=&-\frac{\left( gH\right) ^{2}}{4\pi ^{2}}\int_{0}^{\infty }\frac{%
dt}{t^{2}}\left[ e^{-t}\coth (t)\right] \sum_{p=1}^{\infty }\exp \left[ -%
\frac{\beta ^{2}gHp^{2}}{4t}\right] \cos (p\phi ) \nonumber\\
&&+\frac{\left( -gH\right) ^{2}}{4\pi ^{2}}\int_{0}^{\infty }\frac{dt}{t^{2}}%
e^{-t}\sum_{p=1}^{\infty }\exp \left[ -\frac{\beta ^{2}\left( -gH\right)
p^{2}}{4t}\right] \cos (p\phi )
\end{eqnarray}

\vspace{1pt}The first term can be decomposed using the expansion 
\begin{equation}
\coth (t)=\sum_{k=0}^{\infty }\frac{2^{2k}B_{2k}}{(2k)!}t^{2k-1}=\frac{1}{t}+%
\frac{t}{3}+\sum_{k=2}^{\infty }\frac{2^{2k}B_{2k}}{(2k)!}t^{2k-1} 
\end{equation}
where $B_{n}\,$is the n'th Bernoulli number. We define 
\begin{equation}
f_{0}=-\frac{\left( gH\right) ^{2}}{4\pi ^{2}}\int_{0}^{\infty }\frac{dt}{%
t^{2}}\left[ e^{-t}\frac{1}{t}\right] \sum_{p=1}^{\infty }\exp \left[ -\frac{%
\beta ^{2}gHp^{2}}{4t}\right] \cos (p\phi ) 
\end{equation}
\begin{equation}
f_{1}=-\frac{\left( gH\right) ^{2}}{4\pi ^{2}}\int_{0}^{\infty }\frac{dt}{%
t^{2}}\left[ e^{-t}\frac{t}{3}\right] \sum_{p=1}^{\infty }\exp \left[ -\frac{%
\beta ^{2}gHp^{2}}{4t}\right] \cos (p\phi ) 
\end{equation}
\begin{equation}
f_{k\geq 2}=-\frac{\left( gH\right) ^{2}}{4\pi ^{2}}\int_{0}^{\infty }\frac{%
dt}{t^{2}}\left[ e^{-t}\sum_{k=2}^{\infty }\frac{2^{2k}B_{2k}}{(2k)!}%
t^{2k-1}\right] \sum_{p=1}^{\infty }\exp \left[ -\frac{\beta ^{2}gHp^{2}}{4t}%
\right] \cos (p\phi ). 
\end{equation}
The unstable mode gives a contribution to $U_{\pm }$ which we write as
\begin{equation}
f_{-1}=+\frac{\left( -gH\right) ^{2}}{4\pi ^{2}}\int_{0}^{\infty }\frac{dt}{%
t^{2}}e^{-t}\sum_{p=1}^{\infty }\exp \left[ -\frac{\beta ^{2}(-gH)p^{2}}{4t}%
\right] \cos (p\phi ) 
\end{equation}
using the analytic continuation discussed in section III.
The complete expression for $U_{\pm }$ is the sum
\begin{equation}
U_{\pm }=f_{0}+f_{1}+f_{k\geq 2}+f_{-1}. 
\end{equation}

\vspace{1pt}Using the integral representation of the Bessel function 
\begin{equation}
K_{\nu }(z)=\frac{1}{2}\left( \frac{z}{2}\right) ^{\nu }\int_{0}^{\infty
}dt\,\frac{1}{t^{\nu +1}}\exp \left[ -t-\frac{z^{2}}{4t}\right] 
\end{equation}
we write $f_{0}$ as 
\begin{equation}
f_{0}=-\frac{2gH}{\pi ^{2}\beta ^{2}}\sum_{p=1}^{\infty }\frac{\cos (p\phi )%
}{p^{2}}K_{2}(p\beta \sqrt{gH}),
\end{equation}
and $f_{1}\,$is similarly 
\begin{equation}
f_{1}=-\frac{\left( gH\right) ^{2}}{6\pi ^{2}}\sum_{p=1}^{\infty }\cos
(p\phi )K_{0}(p\beta \sqrt{gH}).
\end{equation}
The $f_{-1}$ term can be done by considering the integral as a function of $%
-gH$, and gives 
\begin{equation}
f_{-1}=-\frac{gH\left( -gH\right) ^{1/2}}{\pi ^{2}\beta }\sum_{p=1}^{\infty }%
\frac{\cos (p\phi )}{p}K_{1}(p\beta \sqrt{-gH}) 
\end{equation}
where the analytic continuation is again specified by $\lim_{\varepsilon
\rightarrow 0}\sqrt{-gH-i\varepsilon }=-i\sqrt{gH}$. These intermediate
forms will require resummation to obtain the high temperature limit.

Resummation of $f_{-1}$, $f_{0}\,$and $f_{1}$ is made possible by a set of
identities for Bessel function sums
which are quite useful in finite temperature field theory
\cite{Meisinger:2001fi}.
For completeness, a brief derivation is given in Appendix B. 
The first of these identitites
is \cite{GandR} 
\begin{equation}
\sum_{p=1}^{\infty }K_{0}(pz)\cos (p\phi )=\frac{1}{2}\left[ \gamma +\ln
\left( \frac{z}{4\pi }\right) \right] +\sum_{l}\,^{^{^{\prime }}}\frac{\pi }{%
2}\left[ \frac{1}{\sqrt{z^{2}+\left( \phi -2l\pi \right) ^{2}}}-\frac{1}{%
2\pi \left| l\right| }\right] 
\end{equation}
where $\gamma $ is Euler's constant. The notation $\sum_{l}\,^{^{^{\prime
}}}\,$indicates that the singular $1/\left| l\right| \,$term is omitted when 
$l=0$. This leads immediately to the formula 
\begin{equation}
f_{1}=-\frac{\left( gH\right) ^{2}}{6\pi ^{2}}\left\{ \frac{1}{2}\left[
\gamma +\ln \left( \frac{\beta \sqrt{gH}}{4\pi }\right) \right]
+\sum_{l}\,^{^{^{\prime }}}\frac{\pi }{2}\left[ \frac{1}{\sqrt{\beta
^{2}gH+\left( \phi -2l\pi \right) ^{2}}}-\frac{1}{2\left| l\right| \pi }%
\right] \right\} \text{.} 
\end{equation}
The remaining Bessel function identities yield 
\begin{eqnarray}
f_{0} &=&-\frac{4}{\pi ^{2}\beta ^{4}}\left[ \frac{\pi ^{4}}{90}-\frac{\pi
^{2}\phi ^{2}}{12}+\frac{\pi \phi ^{3}}{12}-\frac{\phi ^{4}}{48}\right]
\nonumber \\
&&+\frac{gH}{\pi ^{2}\beta ^{2}}\left[ \frac{\pi ^{2}}{6}-\frac{\pi \phi }{2}%
+\frac{\phi ^{2}}{4}\right] 
\nonumber\\
&&-\frac{\left( gH\right) ^{2}}{8\pi ^{2}}\left[ \ln \left( \frac{\beta 
\sqrt{gH}}{4\pi }\right) +\gamma -\frac{3}{4}\right]
\nonumber \\
&&-\frac{1}{\pi \beta ^{4}}\sum_{l}\,^{^{^{\prime }}}%
\Biggl\{ \frac{1}{3} %
\left[ \beta ^{2}gH+\left( \phi -2\pi l\right) ^{2}\right] ^{3/2}-\frac{1}{3}%
\left| \phi -2\pi l\right| ^{3} \nonumber\\
&&\phantom{-\frac{1}{\pi \beta ^{4}}\sum_{l}\,^{^{^{\prime }}}\Biggl\{}
-\frac{1}{2}\left| \phi -2l\pi \right| \beta
^{2}gH-\frac{\beta ^{4}\left( gH\right) ^{2}}{16\pi \left| l\right| 
}\Biggr\}
\end{eqnarray}
and 
\begin{eqnarray}
f_{-1} &=&-\frac{\left( gH\right) ^{2}}{4\pi ^{2}}\left[ \ln \left( \frac{%
\beta \sqrt{-gH}}{4\pi }\right) +\gamma -\frac{1}{2}\right] -\frac{gH}{\pi
^{2}\beta ^{2}}\left[ \frac{1}{4}\phi ^{2}-\frac{\pi }{2}\phi +\frac{\pi ^{2}%
}{6}\right] \nonumber\\
&&+\frac{gH}{2\pi \beta ^{2}}\sum_{l}\,^{^{^{\prime }}}\left[ \sqrt{-\beta
^{2}gH+\left( \phi -2\pi l\right) ^{2}}-\left| \phi -2\pi l\right| +\frac{%
\beta ^{2}gH}{4\pi \left| l\right| }\right] \text{.}
\end{eqnarray}
Although both $f_{0}$ and $f_{-1}$ appear to contain polynomials in $\phi $
which are not manifestly periodic, these terms are the representation on the
range $0\,$to $2\pi \,$of periodic functions. As explained in Appendix B,
they are obtained from the Bernoulli polynomials.

The analytic continuation of the logarithm in $f_{-1}$ gives 
\begin{eqnarray}
f_{-1} &=&-\frac{\left( gH\right) ^{2}}{4\pi ^{2}}\left[ \ln \left( \frac{%
\beta \sqrt{gH}}{4\pi }\right) -i\frac{\pi }{2}+\gamma -\frac{1}{2}\right] -%
\frac{gH}{\pi ^{2}\beta ^{2}}\left[ \frac{1}{4}\phi ^{2}-\frac{\pi }{2}\phi +%
\frac{\pi ^{2}}{6}\right] \nonumber\\
&&+\frac{gH}{2\pi \beta ^{2}}\sum_{l}\,^{^{^{\prime }}}\left[ \sqrt{-\beta
^{2}gH+\left( \phi -2\pi l\right) ^{2}}-\left| \phi -2\pi l\right| +\frac{%
\beta ^{2}gH}{4\pi \left| l\right| }\right] \text{.}
\end{eqnarray}
Note that imaginary terms can potentially arise from a finite
number of the square roots in $f_{-1}$, depending on the value of $\phi $.

\vspace{1pt}The remaining term, $f_{k\geq 2}$, is

\begin{equation}
f_{k\geq 2}\equiv -\frac{\left( gH\right) ^{2}}{8\pi ^{2}}\int_{0}^{\infty }%
\frac{dt}{t^{2}}e^{-t}\sum_{k=2}^{\infty }\frac{2^{2k}B_{2k}}{(2k)!}%
t^{2k-1}\sum_{p\neq 0}\exp \left[ -\frac{p^{2}}{4t}\beta ^{2}gH+ip\phi
\right] \text{.} 
\end{equation}
Once more performing a $\theta _{4}$ transformation we obtain

\begin{equation}
f_{k\geq 2}=\frac{\left( gH\right) ^{3/2}}{4\pi ^{3/2}\beta }%
\sum_{k=2}^{\infty }\frac{2^{2k}B_{2k}}{(2k)!}\int_{0}^{\infty
}dt\,\,t^{2k-5/2}e^{-t}\left\{ \frac{\beta \sqrt{gH}}{\sqrt{4\pi t}}%
-\sum_{l}\exp \left[ -t\frac{1}{\beta ^{2}gH}\left( \phi -2\pi l\right)
^{2}\right] \right\} \text{.} 
\end{equation}
The first term in curly brackets represents a finite contribution to
the renormalization of $(gH)^{2}$ and is evaluated in Appendix C; the
integral over $t\,\,$in the second term can be performed analytically, and
we obtain

\begin{equation}
f_{k\geq 2}=C_{1}\frac{\left( gH\right) ^{2}}{8\pi ^{2}}-\frac{\left(
gH\right) ^{3/2}}{4\pi ^{3/2}\beta }\sum_{k=2}^{\infty }\frac{2^{2k}B_{2k}}{%
(2k)!}\Gamma (2k-3/2)\sum_{l}\frac{\left( \beta ^{2}gH\right) ^{2k-3/2}}{%
\left[ \beta ^{2}gH+\left( \phi -2\pi l\right) ^{2}\right] ^{2k-3/2}}\text{.}
\end{equation}
The numerical value of the constant is approximately $C_{1}=-0.01646$.

The complete expression for $U_{\pm }\,$is 
\begin{eqnarray}
U_{\pm } &=&-\frac{4}{\pi ^{2}\beta ^{4}}\left[ \frac{\pi ^{4}}{90}-\frac{%
\pi ^{2}\phi ^{2}}{12}+\frac{\pi \phi ^{3}}{12}-\frac{\phi ^{4}}{48}\right]
\nonumber\\
&&-\frac{11}{24\pi ^{2}}\left( gH\right) ^{2}\left[ \ln \left( \frac{\beta 
\sqrt{gH}}{4\pi }\right) +\gamma \right] +\frac{7}{32\pi ^{2}}\left(
gH\right) ^{2}+i\frac{1}{8\pi }\left( gH\right) ^{2} \nonumber\\
&&-\frac{1}{\pi \beta ^{4}}\sum_{l}\,^{^{^{\prime }}}\left\{ \frac{1}{3}%
\left[ \beta ^{2}gH+\left( \phi -2\pi l\right) ^{2}\right] ^{3/2}-\frac{1}{3}%
\left| \phi -2\pi l\right| ^{3}-\frac{\beta ^{4}\left( gH\right) ^{2}}{16\pi
\left| l\right| }\right\} \nonumber\\
&&-\frac{\left( gH\right) ^{2}}{12\pi }\sum_{l}\,^{^{^{\prime }}}\left[ 
\frac{1}{\sqrt{\beta ^{2}gH+\left( \phi -2l\pi \right) ^{2}}}-\frac{1}{%
2\left| l\right| \pi }\right] \nonumber\\
&&+\frac{gH}{2\pi \beta ^{2}}\sum_{l}\,^{^{^{\prime }}}\left[ \sqrt{-\beta
^{2}gH+\left( \phi -2\pi l\right) ^{2}}+\frac{\beta ^{2}gH}{4\pi \left|
l\right| }\right] \nonumber\\
&&+C_{1}\frac{\left( gH\right) ^{2}}{8\pi ^{2}}-\frac{\left( gH\right) ^{3/2}%
}{4\pi ^{3/2}\beta }\sum_{k=2}^{\infty }\frac{2^{2k}B_{2k}}{(2k)!}\Gamma
(2k-3/2)\sum_{l}\frac{\left( \beta ^{2}gH\right) ^{2k-3/2}}{\left[ \beta
^{2}gH+\left( \phi -2\pi l\right) ^{2}\right] ^{2k-3/2}}\text{.}
\label{goodequation}
\end{eqnarray}
This can be added to the previous, much simpler, expressions for $V(T=0)$
and $U_{0}$ to give a complete expression for $V$. With the exception of the
first term, the entire expression is manifestly periodic in $\phi $. This
first term is essentially the fourth Bernoulli polynomial, and is valid as
written for the range $0\leq \phi <2\pi $. Note that the order $T^{2}\,$%
terms involving the second Bernoulli polynomial have disappeared from the
final expression, the result of a cancellation of contributions from $f_{0}$
and $f_{-1}$. The omission of the tachyonic mode contribution $f_{-1}$ led
to a spurious $T^{2}$ term in an early calculation of $V(\phi =0)$
\cite{Dittrich:nh}. The $O(T^4)$ term dominates the effective potential
at high temperatures, which implies that $\phi$ is always zero
at the global minimum of the effective potential.

It is possible to extract from $V$ a simple representation for $V_{I}$. Let $%
L_{+}$ be the largest integer such that $2\pi L_{+}<\beta \sqrt{gH}+\phi $
and $L_{-}\,$be the smallest integer such that $2\pi L_{-}>-\beta \sqrt{gH}%
+\phi $. Then $V_{I}$ is given by 
\begin{equation}
V_{I}=-i\frac{gH}{2\pi \beta ^{2}}\sum_{l=L_{-}}^{L_{+}}\left[ \sqrt{\beta
^{2}gH-\left( \phi -2\pi l\right) ^{2}}\right] 
\end{equation}
which represents a generalization of an expression first derived by Cabo 
\textit{et al.} \cite{Cabo:1980js}
for the case $\phi =0$ using different methods. In addition
to the derivation from the high temperature expanion,
we have also derived this result for arbitrary $\phi $
using their methods, and have verified numerically that this expression is
equal to the low-temperature form derived in the previous
section; see reference \cite{Meisinger:1997jt} for
graphs of this function.

As a simple check on our results, we consider the $H\rightarrow 0$
limit. This limit follows quickly from the expressions for $V(T=0)$, $U_{0}$%
, and $U_{\pm }$. We have 
\begin{equation}
V_{R}(H=0)=-6\frac{\pi ^{2}}{90\beta ^{4}}+\frac{4}{\pi ^{2}\beta ^{4}}%
\left[ \frac{\pi ^{2}\phi ^{2}}{12}-\frac{\pi \phi ^{3}}{12}+\frac{\phi ^{4}%
}{48}\right] 
\end{equation}
and of course $V_{I}(H=0)=0$, in agreement with the results of
references \cite{Gross:1981br,Weiss:1981rj,Weiss:1982ev}.
It is invariant under the substitution $\phi
\rightarrow 2\pi -\phi $, reflecting the $Z(2)\,$\ invariance of the gauge
theory. The minimum of $V_{R}$ occurs at $\phi =0$, or equivalently $\phi
=2\pi $, where $Tr_{F}({\mathcal P})=2$. For $\phi =0$, $V_{R}$ is simply the
free energy of a black body with $6\,$degrees of freedom, resulting from $%
3\, $colors in the adjoint representation, each having $2$ spin states. This
is the naive behavior expected at high temperatures: a free gas of gauge
bosons.

The $\phi =0\,$limit is somewhat more complicated than the $H=0$
limit. The complete expression is

\begin{eqnarray}
V &=&\frac{11g^{2}H^{2}}{48\pi ^{2}}\,\,\ln \left( {\frac{gH}{{\mu }_{0}^{2}}%
}\right) -6\frac{\pi ^{2}}{90\beta ^{4}} \nonumber\\
&&-\frac{11}{24\pi ^{2}}\left( gH\right) ^{2}\left[ \ln \left( \frac{\beta 
\sqrt{gH}}{4\pi }\right) +\gamma \right] +\frac{7}{32\pi ^{2}}\left(
gH\right) ^{2} \nonumber\\
&&-\frac{1}{\pi \beta ^{4}}\sum_{l}\,^{^{^{\prime }}}\left\{ \frac{1}{3}%
\left[ \beta ^{2}gH+\left( 2\pi l\right) ^{2}\right] ^{3/2}-\frac{1}{3}%
\left| 2\pi l\right| ^{3}-\frac{\beta ^{4}\left( gH\right) ^{2}}{16\pi
\left| l\right| }\right\} \nonumber\\
&&-\frac{\left( gH\right) ^{2}}{12\pi }\sum_{l}\,^{^{^{\prime }}}\left[ 
\frac{1}{\sqrt{\beta ^{2}gH+\left( 2\pi l\right) ^{2}}}-\frac{1}{2\pi \left|
l\right| }\right] \nonumber\\
&&+\frac{gH}{2\pi \beta ^{2}}\sum_{l}\,^{^{^{\prime }}}\left[ \sqrt{-\beta
^{2}gH+\left( 2\pi l\right) ^{2}}+\frac{\beta ^{2}gH}{4\pi \left| l\right| }%
\right] \nonumber\\
&&+C_{1}\frac{\left( gH\right) ^{2}}{8\pi ^{2}}-\frac{\left( gH\right) ^{3/2}%
}{4\pi ^{3/2}\beta }\sum_{k=2}^{\infty }\frac{2^{2k}B_{2k}}{(2k)!}\Gamma
(2k-3/2)\sum_{l}\frac{\left( \beta ^{2}gH\right) ^{2k-3/2}}{\left[ \beta
^{2}gH+\left( 2\pi l\right) ^{2}\right] ^{2k-3/2}}\text{.}
\end{eqnarray}
Let $L$ be the largest positive integer such that $2\pi L<\beta \sqrt{gH}$.
Then the imaginary part of $V$, $V_{I}$ , is given by 
\begin{equation}
V_{I}=-i\frac{gH}{2\pi \beta ^{2}}\sum_{l=-L}^{L}\left[ \sqrt{\beta
^{2}gH-\left( 2\pi l\right) ^{2}}\right] 
\end{equation}
which is precisely the result of Cabo \textit{et al.}
\cite{Cabo:1980js}.
The high temperature
limit of $V_{R}\,$\ to order $T^{0\,}$is

\begin{eqnarray}
V_{R} &=&\frac{11g^{2}H^{2}}{48\pi ^{2}}\,\,\ln \left( {\frac{gH}{{\mu }%
_{0}^{2}}}\right) -6\frac{\pi ^{2}}{90\beta ^{4}} \nonumber\\
&&-\frac{11}{24\pi ^{2}}\left( gH\right) ^{2}\left[ \ln \left( \frac{\beta 
\sqrt{gH}}{4\pi }\right) +\gamma \right] +\frac{7+4C_{1}}{32\pi ^{2}}\left(
gH\right) ^{2} \nonumber\\
&&-\frac{(gH)^{3/2}}{2\pi \beta }\left[ \frac{5}{6}+\frac{1}{2\pi ^{1/2}}%
\sum_{k=2}^{\infty }\frac{2^{2k}B_{2k}}{(2k)!}\Gamma (2k-3/2)\right]
+O\left( \beta ^{2}\left( gH\right) ^{3}\right)
\end{eqnarray}

\vspace{1pt}The sum in the last term is converted to an integral
and evaluated numerically in Appendix C.
The result is 
\begin{eqnarray}
V_{R} &=&-\frac{11g^{2}H^{2}}{24\pi ^{2}}\,\,\left[ \ln \left( \frac{\beta
\mu _{0}}{4\pi }\right) -\gamma \right] -6\frac{\pi ^{2}}{90\beta ^{4}} 
\nonumber\\
&&+\frac{7+4C_{1}}{32\pi ^{2}}\left( gH\right) ^{2}-C_{2}\frac{(gH)^{3/2}}{%
2\pi \beta }+O\left( \beta ^{2}\left( gH\right) ^{3}\right)
\end{eqnarray}
where $C_{2}$ has the approximate value $C_{2}=0.82778$, in agreement with
the work of Ninomiya and Sakai \cite{Ninomiya:eq}
and of Persson \cite{Persson:1996zy}.
Appendix C also proves that our expression for $C_{2}$ is equivalent to
that given in reference \cite{Ninomiya:eq}.
In the high temperature limit, only the $L=0$ term contributes
to $V_{I}$, which gives 
\begin{equation}
V_{I}=-i\frac{\left( gH\right) ^{3/2}}{2\pi \beta } 
\end{equation}
which also agrees with references \cite{Ninomiya:eq,Persson:1996zy}.

We can identify the coefficient of $\left( gH\right) ^{2}$
in $V_R$ as $%
1/2g_{eff}^{2}(T)$, \textit{i.e.}, 
\begin{equation}
g_{eff}^{2}\left( T\right) =\frac{1}{-\frac{11}{12\pi ^{2}}\,\,\left[ \ln
\left( \frac{\beta \mu _{0}}{4\pi }\right) -\gamma \right] +\frac{7+4C_{1}}{%
16\pi ^{2}}} 
\end{equation}
which is positive because $\beta \mu _{0}<1$ and goes to zero as $%
T\rightarrow \infty $, in accord with asymptotic freedom. The minimum of $%
V_{R}$ occurs at 
\begin{equation}
\beta ^{2}gH\simeq \left[ \frac{36\pi C_{2}}{-44\,\,\left[ \ln \left( \frac{%
\beta \mu _{0}}{4\pi }\right) -\gamma \right] +21+12C_{1}}\right] ^{2}\simeq
\left[ \frac{3C_{2}g_{eff}^{2}\left( T\right) }{4\pi }\right] ^{2}\text{.} 
\end{equation}
The minimum value of this dimensionless variable goes slowly to zero at $T$
goes to infinity. As $T$ ranges from $2\mu _{0}\,$to $10\mu _{0}$, the
minimum decreases from $0.248$ to $0.131$, so the assumption that $\beta
^{2}gH$ is small can be justified at temperatures not much larger than $\mu
_{0}$. Thus perturbation theory requires that the Saviddy model has $H\neq
0\,$at arbitrarily high temperatures, and the standard perturbative state ($%
H=0$) is a local maximum of the effective potential.
This result must be considered
more reliable than the similar zero-temperature result, because asymptotic
freedom applies. However, the free energy of the $H\neq 0\,$state continues
to have a non-zero imaginary part at high $T$, so the Saviddy state
(constant $H\neq 0$) is also unstable. Our results here are in complete
agreement with the earlier work of Ninomiya and Sakai \cite{Ninomiya:eq}.
The deviation of $V_{R}\,$from the black body, ($H=0$%
) result is less than $1\%\,$\ for all temperatures above $2\mu _{0}$, so
any indication of this effect, if it exists, 
is essentially unobservable in lattice
determinations of the pressure and other thermodynamic quantities.

\section{High T Behavior and Dimensional Reduction}

As shown in the previous section, the leading behavior of $V$ 
at high temperature is of order $T^{4}$, and is
independent of $H$. Due to the cancellation of order $T^{2}$ terms, 
the next-to-leading term is of order $T$. The origin
of this $O(T)$ term has been obscure, and was non-trivial to obtain via zeta
function methods even in the case of free fields
\cite{Actor:1986zf,Actor:1987cf}.
In this
section, we show how the $O(T)$ term arises naturally from the 
$n=0$ mode in the context of dimensional reduction
\cite{Ginsparg:1980ef,Appelquist:vg,Nadkarni:1982kb,Nadkarni:1988fh}.
The application of dimensional
reduction is straightforward: the functional determinant can be regarded as
an infinite product of three-dimensional functional integrals in
which each Matsubara frequency $n$ has a mass of $\left| 2\pi n/\beta
\right| $. The $n\neq 0$ modes contribute a $T^{4}$ term which is
independent of $H$, as well as a logarithmic correction term to the
classical action. The $n=0$ contribution must be $O(T)$ because
the only $T$ dependence in this mode arises from the replacement 
\begin{equation}
\int \frac{d^{4}k}{\left( 2\pi \right) ^{4}}\rightarrow \sum_{n}T\int \frac{%
d^{3}k}{\left( 2\pi \right) ^{3}}\text{.} 
\end{equation}
The contribution of the $n=0$ mode to $V$, which we write as $V_{n=0}$\ ,
can be treated by the techniques developed above. After using Schwinger's
proper time representation, $V_{n=0}$ has the form 
\begin{equation}
V_{n=0}={-}\,\sum_{m=0}^{\infty }\,\sum_{\pm }\,{\frac{1}{\beta }}\,\,{\frac{%
gH}{2\pi }}\,\int \,{\frac{dk_{3}}{2\pi }}\int \frac{dt}{t}\,\exp \left\{
-t\left[ \left( {\frac{\phi }{\beta }}\right) ^{2}+2gH\left( m+{\frac{1}{2}}%
\pm 1\right) +k_{3}^{2}\right] \right\} 
\end{equation}
which becomes, after summation over $m$ and integration over $k_{3}$%
\begin{equation}
V_{n=0}=-\,{\frac{gH}{4\pi ^{3/2}\beta }}\int \frac{dt}{t^{3/2}}\,\left[
e^{-tgH}\coth (tgH)+e^{tgH}\right] \exp \left[ -t\left( {\frac{\phi }{\beta }%
}\right) ^{2}\right] \text{.} 
\end{equation}
After expandsion of the $\coth $ and integration over $t$, we find 
\begin{equation}
V_{n=0}=-\,\frac{\left( gH\right) ^{3/2}}{4\pi ^{3/2}\beta }\,\left[
\sum_{k=0}^{\infty }\frac{2^{2k}B_{2k}}{(2k)!}\frac{\Gamma (2k-3/2)(\beta
^{2}gH)^{2k-3/2}}{\left( \phi ^{2}+\beta ^{2}gH\right) ^{2k-3/2}}\right]
-i\,\,{\frac{\left( gH\right) }{2\pi \beta }}^{3/2}\,\,\sqrt{1-\frac{\phi
^{2}}{\beta ^{2}gH}} 
\end{equation}
which may be compared with the similar terms contained in 
Eq. \ref{goodequation}.
The loss of periodicity
in $\phi $ is expected when treating only the $n=0$ mode; similar behavior
has been observed in calculations of the dimensionally reduced theory with $%
H=0$ \cite{Kajantie:1998yc}.
Note that the $Z(2)$ symmetry does remain as the
discrete $\phi \rightarrow -\phi $ symmetry.

Direct comparison with the high temperature expansion for $V\,$\
is simplest when $\phi \,$is taken to be $0$, which
is appropriate at high temperatures. A different form for $V_{n=0}$
can be obtained in this case by integrating first over $k_{3}$ and then over 
$t$, deferring the summation over $m$. This gives 
\begin{eqnarray}
V_{n=0}\left( \phi =0\right) &=&\frac{\left( gH\right) ^{3/2}}{2\pi \beta }%
\,\sum_{m=0}^{\infty }\,\sum_{\pm }\,\,
\left( 2m+{1}\pm 2\right) ^{1/2} \nonumber\\
&=&\,\frac{\left( gH\right) ^{3/2}}{2\pi \beta }\left[ -i-1+2\sqrt{2}\zeta
\left( -1/2,1/2\right) \right] \nonumber\\
&=&\,\frac{\left( gH\right) ^{3/2}}{2\pi \beta }\left[ -i-1+2\sqrt{2}\left( 
\sqrt{2}-1\right) \zeta \left( -1/2\right) \right]
\end{eqnarray}
where the Hurwitz zeta function is defined via the series 
\begin{equation}
\zeta \left( s,\alpha \right) =\sum_{n=0}^{\infty }\frac{1}{(n+\alpha )^{s}}%
\text{.} 
\end{equation}
The real part of this expression has the same form obtained
by Persson \cite{Persson:1996zy}.
The approximate value, after applying a reflection formula for $\zeta \left(
-1/2,1/2\right) $ and summing the resulting series numerically, is 
\begin{equation}
{V}_{n=0}=\frac{\left( gH\right) ^{3/2}}{2\pi \beta }\left[
-i-0.82778\right]. 
\end{equation}
This result for the 
$O(T)$ part of $V$
is in exact agreement with other results 
in the case $\phi = 0$, but here is
completely attributable to the $n=0$ Matsubara mode.
The full mode sum is necessary to recover periodicity in $\phi $.

\section{Magnetic Mass and Stability of $H\neq 0$ at High Temperature}

It has been known for some time \cite{Ninomiya:eq}
that the Savvidy model at high temperature continues to exhibit
the pathologies
associated with its zero-temperature behavior. The real part of the one-loop
effective potential has a minimum at $H\neq 0$, and the imaginary part of
the potential is non-zero at that minimum. This result must be considered
more reliable than the similar zero-temperature result, however,
because asymptotic freedom applies. 
As we will
show below in the context of the Savvidy model, the effect of any gluon
condensate at high temperature would be difficult to observe in lattice
calculations of thermodynamic quantities, because the condensate effects are
of order $g^{6}T^{4}$. 
Nevertheless, it would be troubling if finite
temperature effects did not act to eliminate the gluon condensate at high
temperatures. It is intuitively appealing that the high temperature behavior
of a gluon gas approaches that of a non-interacting relativistic gas. If the
high temperature behavior is fundamentally this simple, the Savvidy
instability must be removed by some mechanism.

The resolution of the high-temperature stability issue is also of interest
on phenomenological grounds. It has been suggested by Enqvist and Olesen 
\cite{Enqvist:1994rm}
that the large-scale non-Abelian magnetic fields in the early universe may
have provided a mechanism for seeding the galactic dynamo. In their work,
they used an approximation to the Saviddy model at high temperature.
However, certain of their assumptions were questionable. In particular, they
assumed that $gH$ would remain near its $T=0$ value at high temperature. As
we have seen, the $\log \left( H\right) $ term responsible for $H\neq 0\,$at
zero temperature is replaced by a $\log \left( T\right) $ term at high
temperature, and the value of $H$ at high temperature need not be
commensurate with the zero temperature value. They also assumed that the
electric screening mass $M_{e}$, which is of order $gT$, plays a role in
overcoming the tachyonic instability. As we discuss below, it is rather the
magnetic screening mass $M_{m}$, believed to be of order $g^{2}T$, which is
relevant. Later work by Elmfors and Persson \cite{Elmfors:1998dr}
used the magnetic mass
rather than the electric mass. However, they also assumed that a
spontaneously generated magnetic field at high temperature would be close to
the zero temperature value. Using a renormalization group argument, they
showed that $gH$ at $T=0$ is always less than $M_{m}^{2}$, and concluded
that a spontaneous magnetic field would be irrelevant at high temperatures.
They therefore considered only the case of an externally imposed field in
detail.

An ambitious analysis of the behavior of $H$ at high temperatures has been
attempted in the recent work of Skalozub and Bordag \cite{Skalozub:1999bf},
which includes the
effect of two-loop and ring diagrams. They assume the magnetic screening
mass originates solely from having $H\neq 0$. Their result for the real part
of the effective potential indicates that a non-zero $H$ is favored at high
temperature, with $gH$ of order $g^{8/3}T^{2}$. However, they also note that
their expression for the imaginary part of the effective potential is
non-zero, again implying that $H\neq 0\,$is unstable. Unfortunately, our
results for the free energy at one loop disagree with theirs in the term
proportional to $T$. There are additional order $T\,\ $terms in our
expressions for $f_{1}$ and $f_{k\geq 2}$ that are responsible for the
difference. Our results are in agreement with the earlier calculations of
Ninomiya and Sakai \cite{Ninomiya:eq}
and of Persson \cite{Persson:1996zy}.

Rather than exploring a specific origin for the magnetic screening mass, we
simply assume that a magnetic screening mass $M_{m}$ of order $g^{2}T$ is
present, and consider the consequences of that assumption. As we will show
below, if the constant of proportionality is sufficiently large, the Savvidy
instability is removed, and $H=0\,\ $is favored at high temperature. The
addition of the magnetic mass requires the replacement 
\begin{equation}
\left( \frac{2\pi n}{\beta }-{\frac{\phi }{\beta }}\right) ^{2}\rightarrow
\left( \frac{2\pi n}{\beta }-{\frac{\phi }{\beta }}\right) ^{2}+M_{m}^{2} 
\end{equation}
in all sums over Matubara modes. We need only consider the impact of $M_{m}$
on the $n=0$ Matsubara mode, because $M_{m}^{2}$ is assumed of order $%
g^{4}T^{2}$, and is therefore negligible compare to the $T^{2}$ term
occuring when $n\neq 0$. The contribution of the $n=0$ mode to the effective
potential is

\begin{equation}
V_{n=0}={-}\,\sum_{m=0}^{\infty }\,\sum_{\pm }\,{\frac{1}{\beta }}\,\,{\frac{%
gH}{2\pi }}\,\int \,{\frac{dk_{3}}{2\pi }}\int \frac{dt}{t}\,\exp \left\{
-t\left[ \left( {\frac{\phi }{\beta }}\right) ^{2}+M_{m}^{2}+2gH\left( m+{%
\frac{1}{2}}\pm 1\right) +k_{3}^{2}\right] \right\} \text{.}
\end{equation}
It is obvious that $\phi ^{2}$ and $\beta ^{2}M_{m}^{2}$ play the same role
in this expression. If we set $\phi =0$, as is appropriate at high
temperature, we have in the case $M_{m}^{2}<gH$ 
\begin{equation}
V_{n=0}=-\,\frac{\left( gH\right) ^{3/2}}{4\pi ^{3/2}\beta }\,\left[
\sum_{k=0}^{\infty }\frac{2^{2k}B_{2k}}{(2k)!}\frac{\Gamma (2k-3/2)(\beta
^{2}gH)^{2k-3/2}}{\left( \beta ^{2}M_{m}^{2}+\beta ^{2}gH\right) ^{2k-3/2}}%
\right] -i\,\,{\frac{\left( gH\right) }{2\pi \beta }}^{3/2}\,\,\sqrt{1-\frac{%
M_{m}^{2}}{gH}}\text{,}
\end{equation}
and 
\begin{equation}
V_{n=0}=-\,\frac{\left( gH\right) ^{3/2}}{4\pi ^{3/2}\beta }\,\left[
\sum_{k=0}^{\infty }\frac{2^{2k}B_{2k}}{(2k)!}\frac{\Gamma (2k-3/2)(\beta
^{2}gH)^{2k-3/2}}{\left( \beta ^{2}M_{m}^{2}+\beta ^{2}gH\right) ^{2k-3/2}}%
\right] +\,\,{\frac{gH}{2\pi \beta }}\,\,\sqrt{M_{m}^{2}-gH}
\end{equation}
for the case $M_{m}^{2}>gH$. Using the techniques developed in Appendix C,
the evaluation of the infinite series can be transformed into the evaluation
of an integral. We find numerically that $V_{n=0}$ can be well-approximated
for all values of the magnetic mass by the $k=0\,\ $term in the sum. This
approximation is worst at $M_{m}=0$, for which the error is still less than $%
20\%\,$; the error falls to about $6\%$ for $M_{m}^{2}=gH$, and approaches
zero for $M_{m}^{2}\gg gH$. Thus, for $M_{m}^{2}<gH$, we write 
\begin{equation}
V_{n=0}\simeq -\,\frac{1}{3\pi \beta }\,\left( M_{m}^{2}+gH\right)
^{3/2}-i\,\,{\frac{\left( gH\right) }{2\pi \beta }}^{3/2}\,\,\sqrt{1-\frac{%
M_{m}^{2}}{gH}}\text{,}
\end{equation}
with a similar expression in the case $M_{m}^{2}>gH$. All other
contributions to $V$ are as before. 
As in our previous, perturbative analysis of the high-temperature
behavior of $V$,
we can drop all terms in the dimensionless
potential $\beta ^{4}V_{R}$ which are of order $\left( \beta ^{2}gH\right)
^{3}$ or higher. Then $\beta ^{4}V_{R}\,\ $can be written as 
\begin{eqnarray}
\beta ^{4}V_{R} &\simeq &-6\frac{\pi ^{2}}{90}+\frac{1}{2g_{eff}\left(
T\right) ^{2}}\left( \beta ^{2}gH\right) ^{2}-\,\frac{1}{3\pi }\,\left(
\beta ^{2}M_{m}^{2}+\beta ^{2}gH\right) ^{3/2}\nonumber\\
&&+\theta \left( M_{m}^{2}-gH\right) {\frac{\beta ^{2}gH}{2\pi }}\,\,\sqrt{%
\beta ^{2}M_{m}^{2}-\beta ^{2}gH}
\end{eqnarray}
where we have again used the effective, temperature dependent, coupling
constant defined by 
\begin{equation}
g_{eff}^{2}\left( T\right) =\frac{1}{-\frac{11}{12\pi ^{2}}\,\,\left[ \ln
\left( \frac{\beta \mu _{0}}{4\pi }\right) -\gamma \right] +\frac{7+4C_{1}}{%
16\pi ^{2}}}\text{.}
\end{equation}

We now assume that $M_{m}$ can be written to leading order as $%
M_{m}=cg_{eff}^{2}T$, and define $x=\left( \beta ^{2}gH\right) /g_{eff}^{4}$%
. We then have 
\begin{equation}
\beta ^{4}V_{R}\simeq -6\frac{\pi ^{2}}{90}+g_{eff}^{6}\left[ \frac{1}{2}%
\,\,x^{2}-\,\frac{1}{3\pi }\,\left( c^{2}+x\right) ^{3/2}+\theta \left(
c^{2}-x\right) {\frac{x}{2\pi }}\,\,\sqrt{c^{2}-x}\right] \text{.}
\end{equation}
Since only the term in square brackets need be minimized, this form
explicitly shows that the minimum value of $gH\,\,$will be of order $%
g_{eff}^{4}T^{2}$, and that the free energy consists of the usual $T^{4}$
blackbody term plus a $g_{eff}^{6}T^{4}$ correction due to chromomagnetic
effects. Of course, this expression does not include those perturbative
higher-loop corrections to the free energy, which start at order $%
g_{eff}^{2}T^{4}$ but are independent of $H$ 
\cite{Kapusta:1989tk},
nor does it include
higher-loop corrections depending on $H$, which start at $g_{eff}^{6}T^{4}$
\cite{Linde:ts}.

For $c>1/\pi \sqrt{2}\simeq 0.225$, the non-trivial minimum of 
\begin{equation}
\frac{1}{2}\,\,x^{2}-\,\frac{1}{3\pi }\,\left( c^{2}+x\right) ^{3/2}+\theta
\left( c^{2}-x\right) {\frac{x}{2\pi }}\,\,\sqrt{c^{2}-x}
\end{equation}
occurs at $x=c^{2}$. This is degenerate with $x=0$ when 
\begin{equation}
c_{crit}=\frac{2}{3\pi }\,\left[ \left( 2\right) ^{3/2}-1\right] \simeq 0.388%
\text{.}
\end{equation}
If $c\,$is larger than $c_{crit}$, then $H=0$ is the minimum of $V_{R}$ and $%
V_{I}$ vanishes. In other words, the one-loop prediction is that $H=0$
provided 
\begin{equation}
M_{m}>c_{crit}g_{eff}^{2}T\simeq 0.388\,\,g_{eff}^{2}\,T\text{.}
\end{equation}
The numerical coefficient is likely to be changed by higher-loop effects,
but indicates that a sufficiently large magnetic mass will lead to $H=0$.

In principle, the magnetic screening mass can be determined from lattice
measurements of the gluon propagator. In Landau gauge, Heller \textit{et al.}
found that $M_{m}$ was well fit by $0.456(6)\,\,g^{2}(T)T$ over a wide
range of temperatures 
\cite{Heller:1997nq}.
In maximal Abelian gauge, Cucchieri \textit{et al.}
determined $M_{m}=1.48(17)\,T$ at $T=2T_{c}\,$, which is equivalent to $%
0.505\,g^{2}\left( T\right) T$ using the one-loop form for the running
coupling constant assumed by the authors \cite{Cucchieri:2000cy}.
Later, more extensive, work by
the same authors \cite{Cucchieri:2001tw}
found a complicated gauge- and volume-dependent
structure in the magnetic propagator at low momentum inconsistent
with a simple pole mass. Further progress in extracting a magnetic gluon
mass from lattice simulations is thus dependent on progress in understanding
the low-momentum structure of the finite-temperature gluon propagator.
Although values for $M_{m}$ obtained from lattice simulation are large
enough to possibly argue against a Savviddy instability at high temperature,
the relatively small difference, \textit{i.e.}, $c_{crit}=0.388$ versus $%
0.456$, combined with many theoretical uncertainties, provide no definitive
resolution of the stability issue. Indeed, the fact that the values are
commensurate may indicate that the generation of the magnetic mass is
intrinsically related to the Saviddy instability in some unknown way.

\section{Conclusions}

The ground state of non-Abelian gauge theories has two related and important
features: confinement and scale symmetry breaking. The Savvidy state,
consisting of quantum fluctuations around a constant chromomagnetic field
at zero temperature,
provides some insight into the nature of scale symmetry breaking. As we have
seen, the analysis of quantum fluctuations around a constant field at finite
temperature in an $SU(2)$ gauge theory allows us to study aspects of the
interplay between scale symmetry breaking and confinement. 

At low temperatures, a complicated behavior emerges from the
one-loop effective potential.
The real part of $V$, $V_{R}$ has minima near integer values $n$ of
the dimensionless variable $\sqrt(gH)/\pi T$. As $T$ goes to $0$,
the global minimum corresponds to higher values of $n$, and $H$ approaches
its zero temperature value.
The preferred
value of $\phi $ alternates discontinuously between $0\,$and $\pi $,
representing an increasingly rapid set of transitions
between confined and
deconfined phases as the temperature approaches zero.
However, the imaginary part of the effective potential, $V_{I}\,\ $%
never vanishes at the global minimum of $V_{R}$. 
Therefore the Savvidy model
is unstable at low temperatures. 
While there is no
reason to trust a one loop perturbative calculation at low
temperatures, these results demonstrate that gluon propagation
in a non-trivial background can lead to confinement at low temperatures.

At sufficiently high temperatures, the leading term in the free
energy, which is proportional to $T^{4}$, demands that $\phi =0$. The
leading behavior of $V_{R}$ is that of a free gas of massless,
non-interacting gluons. The dominant subleading contribution to $V_{R}$
at one loop
comes from the $n=0$ Matsubara frequency. This term leads to  $gH$
being of order $g^4 T^2$, in turn making a contribution of order
$g^6 T^4$ to the free energy.
The imaginary part of $V$ remains non-vanishing, and thus a constant
chromomagnetic field is unstable in perturbation theory at high
temperatures. 
However, a magnetic mass of the form $M_m = cg^2T$ would naturally
alter the one-loop results in such a way that $\phi =0$, $H=0\,\ $is favored
in the high temperature limit, provided that the coefficient $c$ is
sufficiently large.
This would restore the picture of the high
temperature state as a plasma of weakly interacting gluons.
In this case, the magnetic sector would still contribute
to $V$ at order $g^6T^4$.

The existence of a sufficiently large magnetic mass
is consistent with our knowledge of the magnetic sector
gleaned from the work of Karabali and Nair
on three-dimensional gauge theories in the Hamiltonian formalism
\cite{Karabali:1995ps,Karabali:1996je,Karabali:1996iu,Nair:2002yg}.
They are able to extract a three-dimensional gluon mass,
corresponding to a magnetic mass in four dimensions.
Their estimates of this mass are consistent with those
obtained from lattice simulation.
As we have seen, the masses obtained in the case of $SU(2)$ are not
sufficiently large to confidently assert that $H=0$ is stable.
It would be very useful to incorporate the results of 
Nair {\it et al.} into an effective potential for
the finite temperature, four-dimensional theory.

Over the years, many kinds of field configurations have been
suggested as being responsible for confinement. A constant chromomagnetic
field is the simplest non-trivial field configuration for which the
associated functional determinant can be obtained analytically. It is
interesting to speculate as to what features of this model might carry over
to more complicated field configurations, and how the Polyakov loop might be
driven to confining behavior. The Savvidy state does appear to distinguish
at one loop between the low temperature regime, where $\left\langle
Tr_{F}{\mathcal P}\right\rangle $ is a sensitive function of the temperature,
and the
high temperature regime, where $\left\langle Tr_{F}{\mathcal P}\right\rangle $
goes to
its maximal values of $\pm 2$ in $SU(2)$. Perhaps this behavior is some
distant relative of the deconfinement transition.

Recent work
on phenomenological models of the deconfinement transition
may provide some direction for further theoretical investigation.
In collaboration with Travis Miller, we have constructed two
models for the free energy which give rise to
confinement at low temperatures
\cite{Meisinger:2001cq}.
These models confine by adding a phenomenological
non-perturbative term to the free energy which depends on the Polyakov loop
eigenvalues. 
Both models account very well for many
features of the deconfinement transition observed in lattice simulations. 
A similar approach has also been developed by Dumitru and Pisarski
\cite{Dumitru:2001bf,Dumitru:2001xa,Pisarski:2002ji}.
Based on their work, Sannino has recently proposed 
a phenomenological effective potential for finite temperature QCD
in which the Polyakov loop is coupled to the scalar glueball field
\cite{Sannino:2002wb}.
Because the operator $Tr\,F_{ij}^2$ couples to the scalar glueball sector,
our expression for the Savvidy effective potential shares certain
features with this more phenomenological approach when $H^2$ is identified
as the glueball field expectation value.
In either approach, however, it does not appear that the glueball
field is driving the deconfinement transition; some other mechanism
is still required.

A complete field-theoretic description of
confinement, perhaps using other observables in a role similar to $H$ in the
Savvidy model, should produce an effective action for the Polyakov loop which
yields confinement at low temperature in a natural way.

\appendix

\section{Zero Temperature}

\vspace{1pt}In this appendix, we derive the zero
temperature part of the effective potential, 
following the approach of reference \cite{Nielsen:1978rm}.
The only $H$-dependent part is $V_{\pm }^{(1)}(T=0)$
which we can write as 
\begin{equation}
V_{\pm }^{(1)}(T=0)=\sum_{m=0}^{\infty }\sum_{\pm }\,\,\frac{gH}{2\pi }\mu %
^{\varepsilon }\,\int \,\frac{d^{1-\varepsilon }k}{\left( 2\pi \right)
^{1-\varepsilon }}\left[ \,k^{2}+2gH\left( m+{\frac{1}{2}}\pm 1\right)
-i\delta \right] ^{1/2} 
\end{equation}
after analytic continuation of the $k\,$\ integral to $1-\varepsilon \,$%
dimensions. We have introduced an $i\delta \,\ $to correctly treat the
imaginary part. This introduces the scale parameter $\mu $. Performing the $%
k\,$integration, we obtain 
\begin{eqnarray}
V_{\pm }^{(1)}(T =0)&=&\frac{gH}{2\pi }\mu ^{\varepsilon
}\sum_{m=0}^{\infty }\sum_{\pm }\,\,\frac{1}{(16\pi ^{2})^{(1-\varepsilon
)/4}}\frac{\Gamma (-1+\frac{\varepsilon }{2})}{\Gamma (-\frac{1}{2})\left[
\,2gH\left( m+{\frac{1}{2}}\pm 1\right) -i\delta \right] ^{-1+\frac{%
\varepsilon }{2}}} \nonumber\\
&=&\frac{gH}{2\pi }\mu^{\varepsilon }
\frac{\Gamma (-1+\frac{\varepsilon }{%
2})
\left[ \zeta \left( -1+\frac{\varepsilon 
}{2},\frac{3}{2}-i\delta \right) +\zeta \left( -1+\frac{\varepsilon }{2},-%
\frac{1}{2}-i\delta \right) \right] }
{\Gamma (-\frac{1}{2})\left[ \,2gH\right] ^{-1+\frac{\varepsilon }{2}%
}(16\pi ^{2})^{(1-\varepsilon )/4}}
\end{eqnarray}
where the generalized Riemann function $\zeta (s,a)$, or Hurwitz function is
defined by 
\begin{equation}
\zeta (s,a)=\sum_{n=0}^{\infty }\frac{1}{\left( n+a\right) ^{s}}\text{.} 
\end{equation}
The term in square brackets can be written as 
\begin{eqnarray}
\zeta \left( -1+\frac{\varepsilon }{2},\frac{3}{2}-i\delta \right) &+&\zeta
\left( -1+\frac{\varepsilon }{2},-\frac{1}{2}-i\delta \right) =\nonumber\\
&&2\left( 2^{-1+%
\frac{\varepsilon }{2}}-1\right) \zeta (-1+\frac{\varepsilon }{2})-2^{-1+%
\frac{\varepsilon }{2}}+(-2+i\delta )^{-1+\frac{\varepsilon }{2}}\text{.} 
\end{eqnarray}
We may evaluate $\zeta (-1+\frac{\varepsilon }{2})\,$ to order\ $\varepsilon
\,$using formulas from Actor \cite{Actor:1986zf,Actor:1987cf}.
To order $\varepsilon $, 
\begin{equation}
\zeta \left( -1+\frac{\varepsilon }{2},\frac{3}{2}-i\delta \right) +\zeta
\left( -1+\frac{\varepsilon }{2},-\frac{1}{2}-i\delta \right) \simeq -\frac{%
11}{12}-i\pi \frac{\varepsilon }{4}+C_{0}\varepsilon 
\end{equation}
where $C_{0}$ is a real, finite constant. Using 
\begin{equation}
\Gamma (-1+\frac{\varepsilon }{2})\approx -\frac{2}{\varepsilon }+\gamma -1 
\end{equation}
we find 
\begin{eqnarray}
V_{\pm }^{(1)}(T =0)&=&-{\frac{(gH)^{2}}{4\pi ^{2}}}\left( \frac{2\pi {\mu }%
^{2}}{gH}\right) ^{\varepsilon /2}\left( -\frac{2}{\varepsilon }+\gamma
-1\right) \left[ -\frac{11}{12}-i\pi \frac{\varepsilon }{4}+C_{0}\varepsilon
\right] \nonumber\\
&=&-{\frac{11(gH)^{2}}{24\pi ^{2}\varepsilon }}+{\frac{11(gH)^{2}}{48\pi ^{2}%
}\ln }\left( \frac{gH}{\mu ^{2}}\right) -i\frac{(gH)^{2}}{8\pi }%
+C_{0}^{\prime }(gH)^{2}+O\left( \varepsilon \right) \text{.}
\end{eqnarray}
Renormalization removes the $1/\varepsilon $ pole and
the
renormalization coupling $g(\mu )$ is defined in the standard way. The
complete zero-temperature effective potential is 
\begin{eqnarray}
V(T =0)&=&V^{(0)}+V^{(1)}(T=0) \nonumber\\
&=&\frac{1}{2}H^{2}+{\frac{11(gH)^{2}}{48\pi ^{2}}\ln }\left( \frac{gH}{\mu
^{2}}\right) -i\frac{(gH)^{2}}{8\pi }+C_{0}^{\prime }(gH)^{2}\text{.}
\end{eqnarray}
We are free to introduce a new renormalization group invariant $\mu _{0}$
such that 
\begin{equation}
\mu _{0}=\mu \exp \left[ -\frac{12\pi ^{2}}{11g^{2}}\left(
1+2g^{2}C_{0}^{\prime }\right) \right] 
\end{equation}
yielding finally

\begin{equation}
V(T=0)=\frac{11g^{2}H^{2}}{48\pi ^{2}}\,\,\ln \left( {\frac{gH}{{\mu }%
_{0}^{2}}}\right) -i\frac{(gH)^{2}}{8\pi ^{2}}\text{.} 
\end{equation}

\vspace{1pt}

\section{Identities for Bessel Function Sums}

In this appendix, we prove the Bessel function identities:
\begin{equation}
\sum_{m=1}^{\infty }K_{0}(mr)\cos (m\phi )=\frac{1}{2}\left[ \ln \left( 
\frac{r}{4\pi }\right) +\gamma \right] +\frac{\pi }{2}\sum_{l}\,^{^{^{\prime
}}}\left[ \frac{1}{\sqrt{r^{2}+\left( \phi -2\pi l\right) ^{2}}}-\frac{1}{%
2\pi \left| l\right| }\right]
\end{equation}
\begin{eqnarray}
\sum_{p=1}^{\infty }\frac{1}{p}K_{1}(pz)\cos (p\phi ) &=&-\frac{1}{4}z\left[
\ln \left( \frac{z}{4\pi }\right) +\gamma -\frac{1}{2}\right] +\frac{1}{z}%
\left[ \frac{1}{4}\phi ^{2}-\frac{\pi }{2}\phi +\frac{\pi ^{2}}{6}\right]
\nonumber \\
&&-\frac{\pi }{2z}\sum_{l}\,^{^{^{\prime }}}\left[ \sqrt{z^{2}+\left( \phi
-2\pi l\right) ^{2}}-\left| \phi -2\pi l\right| -\frac{z^{2}}{4\pi \left|
l\right| }\right]
\end{eqnarray}

\begin{eqnarray}
\sum_{p=1}^{\infty }\frac{1}{p^{2}}K_{2}(pz)\cos (p\phi ) &=&\frac{1}{16}%
z^{2}\left[ \ln \left( \frac{z}{4\pi }\right) +\gamma -\frac{3}{4}\right] -%
\frac{1}{2}\left[ \frac{1}{4}\phi ^{2}-\frac{\pi }{2}\phi +\frac{\pi ^{2}}{6}%
\right] \nonumber\\
&&+\frac{2}{z^{2}}\left[ \frac{-1}{48}\phi ^{4}+\frac{\pi }{12}\phi ^{3}-%
\frac{\pi ^{2}}{12}\phi ^{2}+\frac{\pi ^{4}}{90}\right] \nonumber\\
&&+\frac{\pi }{2z^{2}}\sum_{l}\,^{^{^{\prime }}}\Biggl\{ \frac{1}{3}\left[
z^{2}+\left( \phi -2\pi l\right) ^{2}\right] ^{3/2}-\frac{1}{3}\left| \phi
-2\pi l\right| ^{3}\nonumber\\
&&\phantom{+\frac{\pi }{2z^{2}}\sum_{l}\,^{^{^{\prime }}}}\,
-\frac{1}{2}\left| \phi -2l\pi \right| z^{2}-\frac{z^{4}}{%
16\pi \left| l\right| }\Biggr\} \text{.}
\end{eqnarray}
The right-hand sides of the second two equations are 
valid for $\phi$ in the range
$0 \leq \phi < 2 \pi $.
They can be extended to all real values if $%
\phi \,$is replaced by $\left| \phi \right| {\it \ mod}\ 2\pi $ 
on the right hand side of the equation. 
The second and third identitites follow from the first
\cite{Meisinger:2001fi}.

Using a standard integral representation \cite{GandR}
\begin{equation}
K_{0}(pz)=\int_{0}^{\infty }dt\,\frac{\cos (pt)}{\sqrt{t^{2}+z^{2}}}\text{,}
\end{equation}
we can write the left-hand side of the first identity as
\begin{equation}
\sum_{m=1}^{\infty }\cos (m\phi )K_{0}(mr)
=\frac{1}{4\pi }\sum_{m\neq 0}\int dk_{x}dk_{y}\frac{1}{%
k_{x}^{2}+k_{y}^{2}+m^{2}}e^{ik_{x}r+im\phi }.
\end{equation}
We introduce a regulating mass $\mu $, which will be taken to zero at the
end of the calculation, and add and subtract the divergent $m=0$ term,
obtaining 
\begin{eqnarray}
\frac{1}{4\pi }\sum_{m}\int dk_{x}dk_{y}\frac{1}{%
k_{x}^{2}+k_{y}^{2}+m^{2}+\mu ^{2}}e^{ik_{x}r+im\phi }-\frac{1}{4\pi }\int
dk_{x}dk_{y}\frac{1}{k_{x}^{2}+k_{y}^{2}+\mu ^{2}}e^{ik_{x}r}
\end{eqnarray}
which can be written as
\begin{eqnarray}
\frac{1}{4\pi }\sum_{m}\int dk_{x}dk_{y}dk_{z}
\frac{\delta \left( k_{z}-m\right)}{%
k_{x}^{2}+k_{y}^{2}+k_{z}^{2}+\mu ^{2}}
e^{ik_{x}r+ik_{z}\phi }-\frac{1}{4\pi }\int dk_{x}dk_{y}\frac{1}{%
k_{x}^{2}+k_{y}^{2}+\mu ^{2}}e^{ik_{x}r}.
\end{eqnarray}
Using the Poisson summation technique in the form $\sum_{m}\delta \left(
k_{z}-m\right) =\sum_{n}\exp \left( -2\pi ink_{z}\right) $, we obtain 
\begin{equation}
\frac{\pi }{2}\sum_{n}\int \frac{d^{3}k}{\left( 2\pi \right) ^{3}}\frac{4\pi 
}{k^{2}+\mu ^{2}}e^{ik_{x}r+ik_{z}\left( \phi -2\pi n\right) }-\frac{1}{4\pi 
}\int d^{2}k\frac{1}{k^{2}+\mu ^{2}}e^{ik_{x}r}
\end{equation}
The first integral gives a sum of screened Coulomb, or Yukawa, potentials 
\begin{equation}
\frac{\pi }{2}\sum_{n}\left[ \frac{e^{-\mu \sqrt{r^{2}+\left( \phi -2\pi
n\right) ^{2}}}}{\sqrt{r^{2}+\left( \phi -2\pi n\right) ^{2}}}\right] -\frac{%
1}{4\pi }\int d^{2}k\frac{1}{k^{2}+\mu ^{2}}e^{ik_{x}r}
\end{equation}
and both terms appear to be problematic as $\mu \rightarrow 0$. The first
term can be made finite in this limit by subtracting the contribution at $%
r=\phi =0$ for $n\neq 0$. Using the notation $\sum_{n}{}^{^{^{\prime }}}$ to
denote a summation over all $n$ with the omission of the singular term when $%
n=0$, we have 
\begin{equation}
\frac{\pi }{2}\sum_{n}\,^{^{^{\prime }}}\left[ \frac{e^{-\mu \sqrt{%
r^{2}+\left( \phi -2\pi n\right) ^{2}}}}{\sqrt{r^{2}+\left( \phi -2\pi
n\right) ^{2}}}-\frac{e^{-\mu 2\pi \left| n\right| }}{2\pi \left| n\right| }%
\right] +\left[ \frac{\pi }{2}\sum_{n}\,^{^{^{\prime }}}\frac{e^{-\mu 2\pi
\left| n\right| }}{2\pi \left| n\right| }-\frac{1}{4\pi }\int d^{2}k\frac{1}{%
k^{2}+\mu ^{2}}e^{ik_{x}r}\right] \text{.}
\end{equation}
The first sum over $n$ is finite as $\mu \rightarrow 0$,
The second sum over $n$ can be explicitly summed and the final integral
carried out, leading to
\begin{equation}
\frac{\pi }{2}\sum_{n}\,^{^{^{\prime }}}\left[ \frac{e^{-\mu \sqrt{%
r^{2}+\left( \phi -2\pi n\right) ^{2}}}}{\sqrt{r^{2}+\left( \phi -2\pi
n\right) ^{2}}}-\frac{e^{-\mu 2\pi \left| n\right| }}{2\pi \left| n\right| }%
\right] +\left[ -\frac{1}{2}\ln \left[ 1-e^{-2\pi \mu }\right] -\frac{1}{2}%
K_{0}\left( \mu r\right) \right] \text{.}
\end{equation}
In the limit $\mu \rightarrow 0$, we finally obtain
\begin{equation}
\sum_{m=1}^{\infty }K_{0}(mr)\cos (m\phi )=\frac{\pi }{2}\sum_{n}\,^{^{^{%
\prime }}}\left[ \frac{1}{\sqrt{r^{2}+\left( \phi -2\pi n\right) ^{2}}}-%
\frac{1}{2\pi \left| n\right| }\right] +\frac{1}{2}\ln \left( \frac{r}{4\pi }%
\right) +\frac{1}{2}\gamma \text{.}
\end{equation}

The second identity is obtained from the first using the identity
\begin{equation}
\frac{d}{dz}K_{\nu }(z)=-K_{\nu -1}(z)-\frac{\nu }{z}K_{\nu }(z)\text{.} 
\end{equation}
It follows immediately that 
\begin{equation}
\frac{d}{dz}\sum_{p=1}^{\infty }\frac{z}{p}K_{1}(pz)\cos (p\phi
)=-z\sum_{p=1}^{\infty }K_{0}(pz)\cos (p\phi ) 
\end{equation}
so that 
\begin{equation}
\sum_{p=1}^{\infty }\frac{1}{p}K_{1}(pz)\cos (p\phi )=-\frac{1}{z}\int
dz\,\,\,z\sum_{p=1}^{\infty }K_{0}(pz)\cos (pz)+\frac{C(\phi )}{z} 
\end{equation}
where $C(\phi )\,$is an unknown function to be determined later. Integration
yields 
\begin{eqnarray}
\sum_{p=1}^{\infty }\frac{1}{p}K_{1}(pz)\cos (p\phi )=&-&\frac{1}{4}z\left[
\ln \left( \frac{z}{4\pi }\right) +\gamma -\frac{1}{2}\right] \nonumber\\
&-&\frac{\pi }{2z%
}\sum_{l}\,^{^{^{\prime }}}\left[ \sqrt{z^{2}+\left( \phi -2\pi l\right) ^{2}%
}-\frac{z^{2}}{4\pi \left| l\right| }\right] +\frac{C(\phi )}{z}\text{.} 
\end{eqnarray}
The function $C(\phi )\,$is determined using the behavior of $K_{\nu}(z)$ 
for $ z\rightarrow 0$%
\begin{equation}
K_{\nu }(z)\sim \frac{1}{2}\Gamma (\nu )\left( \frac{2}{z}\right) ^{\nu
} 
\end{equation}
and the standard result 
\begin{equation}
\sum_{p=1}^{\infty }\frac{\cos (p\phi )}{p^{2}}=\frac{1}{4}\phi ^{2}-\frac{%
\pi }{2}\phi +\frac{\pi ^{2}}{6} 
\end{equation}
valid for $0\leq \phi <2\pi $. It can be extended to all real values if $%
\phi \,$is replacd by $\left| \phi \right| \ {\it mod} \  2\pi $ on the right hand side of the
equation. 
This implies the leading behavior of the sum as $z\rightarrow 0\,$%
is 
\begin{equation}
\sum_{p=1}^{\infty }\frac{1}{p}K_{1}(pz)\cos (p\phi )\sim \frac{1}{z}%
\sum_{p=1}^{\infty }\frac{\cos (p\phi )}{p^{2}}=\frac{1}{z}\left[ \frac{1}{4}%
\phi ^{2}-\frac{\pi }{2}\phi +\frac{\pi ^{2}}{6}\right] 
\end{equation}
giving us finally 
\begin{eqnarray}
\sum_{p=1}^{\infty }\frac{1}{p}K_{1}(pz)\cos (p\phi ) &=&-\frac{1}{4}z\left[
\ln \left( \frac{z}{4\pi }\right) +\gamma -\frac{1}{2}\right] +\frac{1}{z}%
\left[ \frac{1}{4}\phi ^{2}-\frac{\pi }{2}\phi +\frac{\pi ^{2}}{6}\right]
\nonumber \\
&&-\frac{\pi }{2z}\sum_{l}\,^{^{^{\prime }}}\left[ \sqrt{z^{2}+\left( \phi
-2\pi l\right) ^{2}}-\left| \phi -2\pi l\right| -\frac{z^{2}}{4\pi \left|
l\right| }\right] \text{.}
\end{eqnarray}

The third identity is obtained from the second by a similar argument,
combined with the identity
\begin{equation}
\sum_{p=1}^{\infty }\frac{\cos (p\phi )}{p^{4}}=\frac{-1}{48}\phi ^{4}+\frac{%
\pi }{12}\phi ^{3}-\frac{\pi ^{2}}{12}\phi ^{2}+\frac{\pi ^{4}}{90}\text{,} 
\end{equation} also valid for $0\leq \phi <2\pi $.

\section{Sums involving Bernoulli numbers}

In this appendix, we show how to evaluate the constants
$C_1$ and $C_2$ introduced in section IV.
Using the generating function for the Bernoulli numbers 
\begin{equation}
\frac{t}{e^{t}-1}=\sum_{k=0}^{\infty }\frac{1}{k!}B_{k}t^{k}=1-\frac{1}{2}t+%
\frac{1}{12}t^{2}+\sum_{k=2}^{\infty }\frac{1}{(2k)!}B_{2k}t^{2k} 
\end{equation}
we can write

\begin{eqnarray}
C_1&=&\sum_{k=2}^{\infty }\frac{%
2^{2k}B_{2k}}{(2k)!}\int_{0}^{\infty }dt\,\,t^{2k-3}e^{-t} \nonumber\\
&=&\int_{0}^{\infty }dt\,e^{-t}\frac{1%
}{t^{3}}\sum_{k=2}^{\infty }\frac{2^{2k}B_{2k}}{(2k)!}\,\,t^{2k} \nonumber\\
&=&\int_{0}^{\infty }dt\,e^{-t}\frac{1%
}{t^{3}}\left[ \frac{2t}{e^{2t}-1}-1+t-\frac{1}{3}t^{2}\right]
\end{eqnarray}
which is convergent, and equal numerically to 
\begin{equation}
\frac{\left( gH\right) ^{2}}{8\pi ^{2}}\left( -0.01646\right) 
\end{equation}

\vspace{1pt}Using the same method, we can also evaluate 
\begin{eqnarray}
C_{2} &=&\frac{5}{6}+\frac{1}{2\pi ^{1/2}}\sum_{k=2}^{\infty }\frac{%
2^{2k}B_{2k}}{(2k)!}\Gamma (2k-3/2) \\
&=&\frac{5}{6}+\frac{1}{\sqrt{4\pi }}\int_{0}^{\infty
}dte^{-t}t^{-5/2}\left[ \frac{2t}{e^{2t}-1}
-1+t-\frac{1}{3}t^{2}\right] \nonumber\\
&\approx &0.82778
\end{eqnarray}
Ninomiya and Sakai give an alternative expression for this constant 
\begin{equation}
C_{2}=-1+\frac{1}{\sqrt{4\pi }}\int_{0}^{\infty }dt\,t^{-5/2}\left[ \frac{%
2t\,e^{-3t}}{1-e^{-2t}}-1+2t\right] \text{.} 
\end{equation}
The difference between the two expressions is 
\begin{equation}
\frac{1}{\sqrt{4\pi }}\int_{0}^{\infty }dt\,t^{-5/2}\left[ e^{-t}\left( -1+t+%
\frac{10}{3}t^{2}\right) +1-2t\right] \text{,} 
\end{equation}
the individual terms of which are very badly behaved. Writing this
expression as 
\begin{equation}
\lim_{\varepsilon \rightarrow 0}\frac{1}{\sqrt{4\pi }}\int_{\varepsilon
}^{\infty }dt\,t^{-5/2}\left[ e^{-t}\left( -1+t+\frac{10}{3}t^{2}\right)
+1-2t\right] \text{,} 
\end{equation}
one can show that the difference is zero using integration by parts.

\begin{figure}
\includegraphics[width=5in]{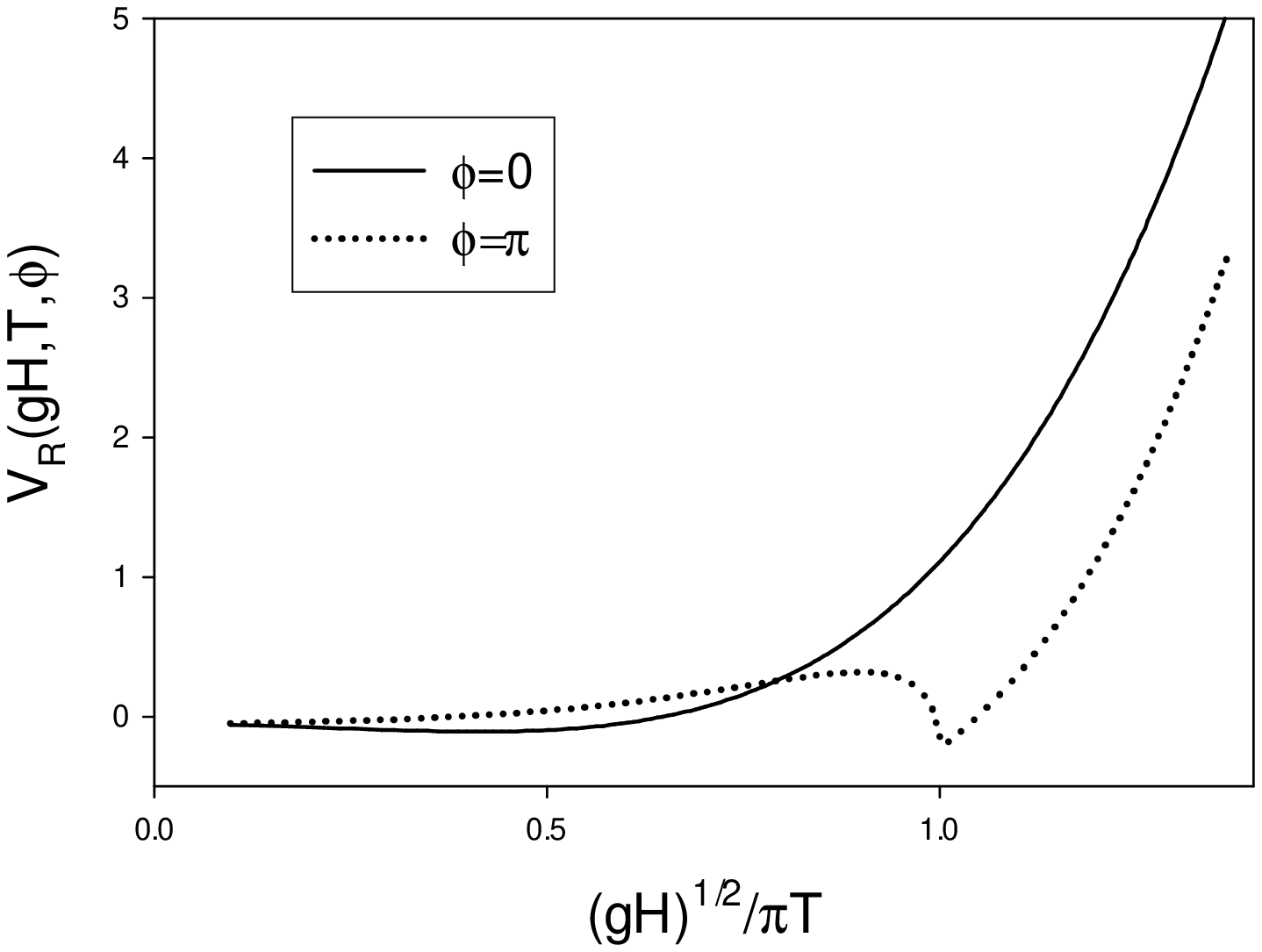}
\caption{$V_R$ versus the scaled variable $\sqrt{gH}/\pi T$ for $T=0.7\mu$}
\label{t=0.7}
\end{figure}  

\begin{figure}
\includegraphics[width=5in]{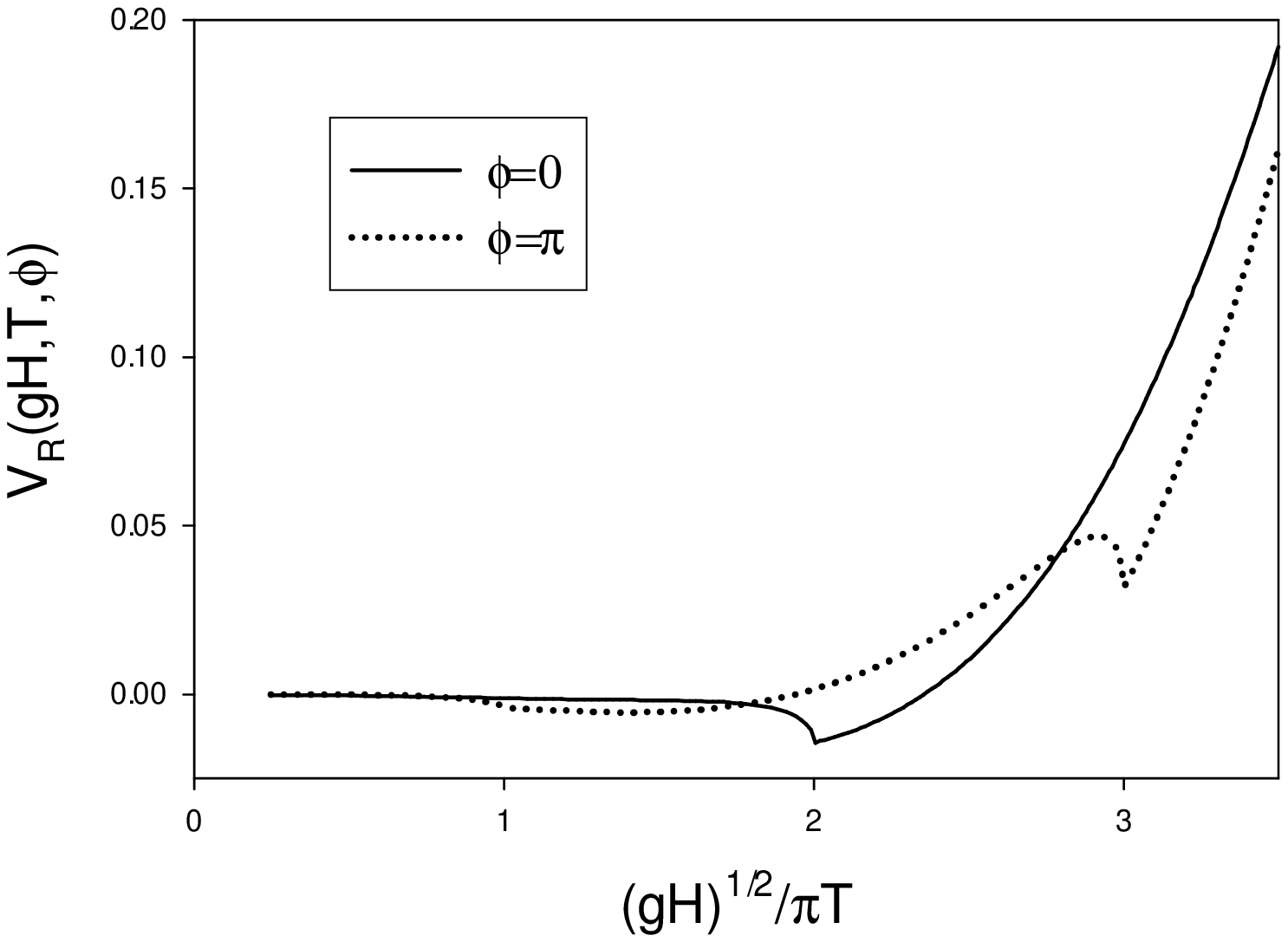}
\caption{$V_R$ versus the scaled variable $\sqrt{gH}/\pi T$ for $T=0.15\mu$}
\label{t=0.15}
\end{figure}  

\begin{figure}
\includegraphics[width=5in]{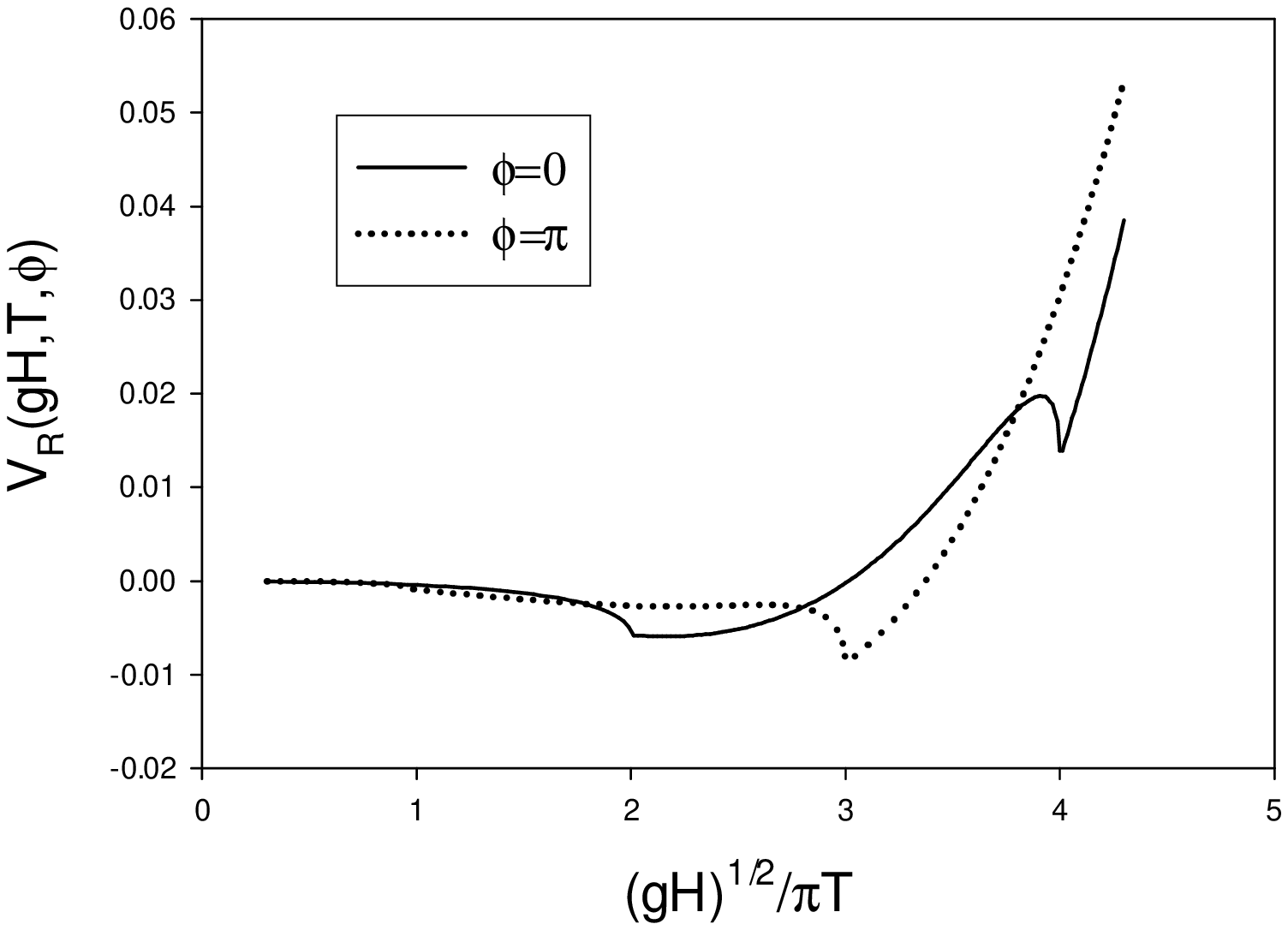}
\caption{$V_R$ versus the scaled variable $\sqrt{gH}/\pi T$ for $T=0.1\mu$}
\label{t=0.1}
\end{figure}  

\begin{figure}
\includegraphics[width=5in]{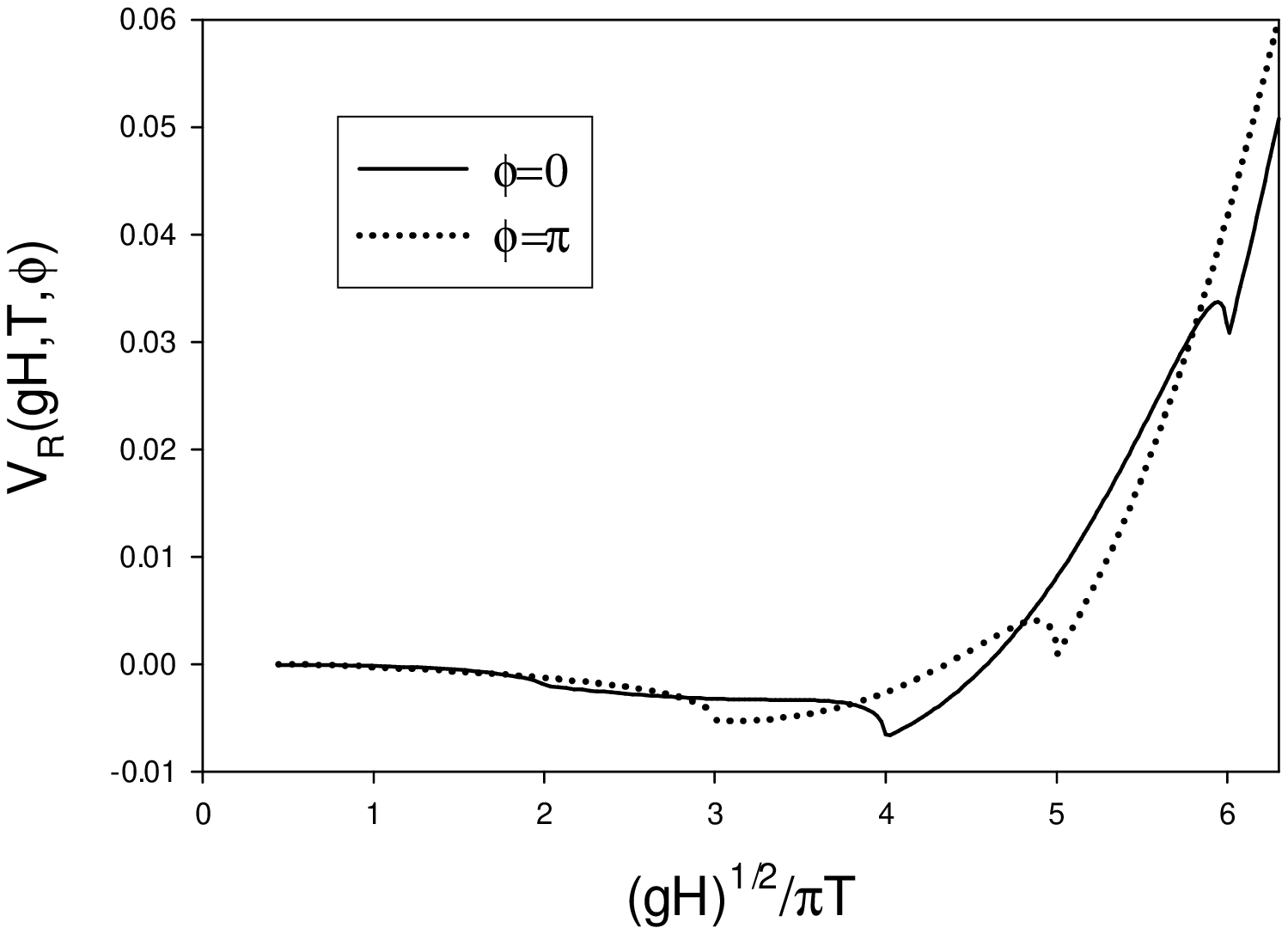}
\caption{$V_R$ versus the scaled variable $\sqrt{gH}/\pi T$ for $T=0.07\mu$}
\label{t=0.07}
\end{figure}  

\begin{figure}
\includegraphics[width=5in]{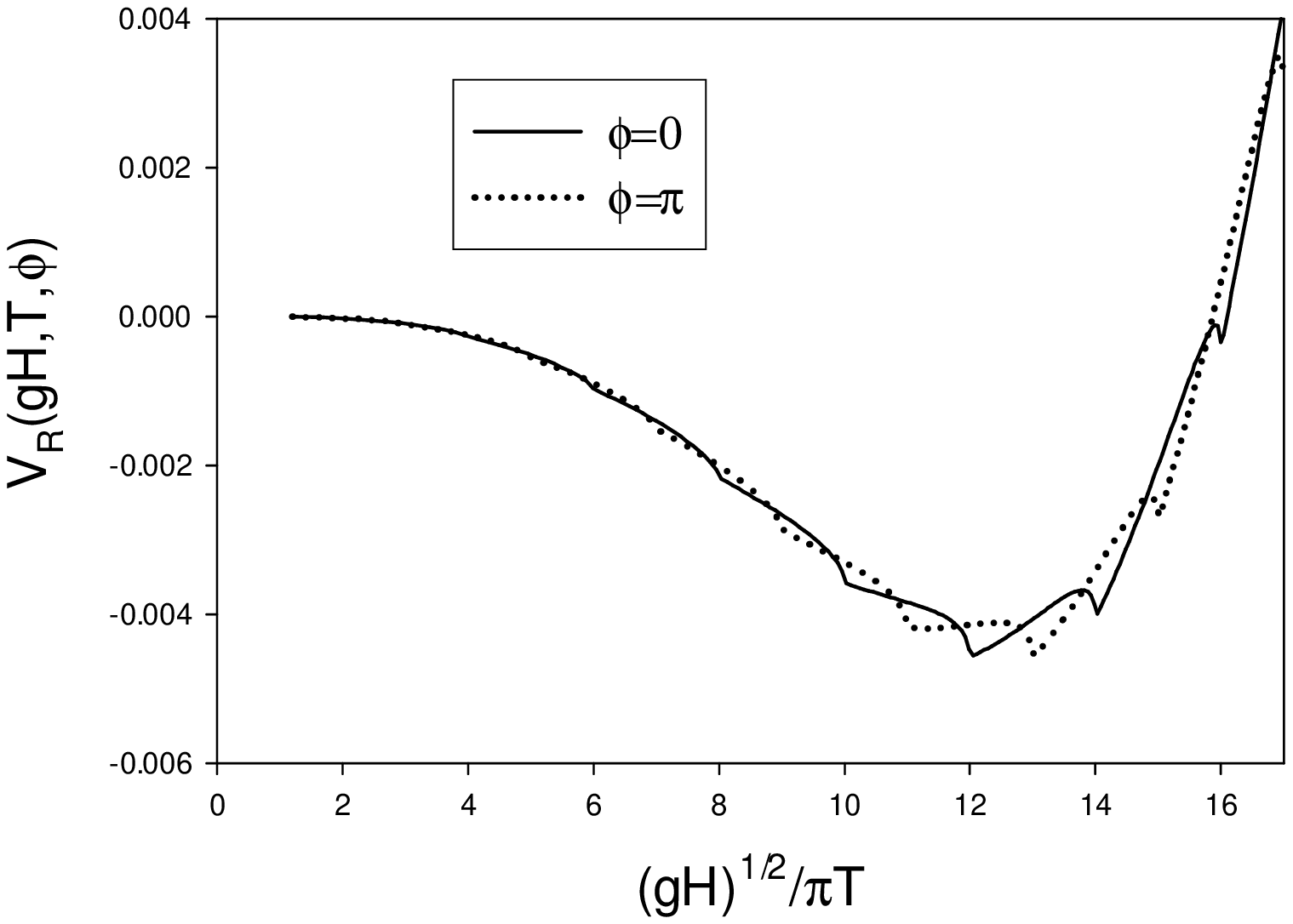}
\caption{$V_R$ versus the scaled variable $\sqrt{gH}/\pi T$ for $T=0.02\mu$}
\label{t=0.02}
\end{figure}  

\end{document}